\documentclass[modern]{aastex62}

\graphicspath{{./}{figures/}}

\submitjournal{PASP}

\shorttitle{Calibrating Iodine Cells for PRVs}
\shortauthors{Wang et al.}

\usepackage{amsmath,amssymb}
\usepackage{xspace}
\usepackage{graphicx}
\usepackage{epstopdf}
\usepackage{graphicx}
\usepackage{epsfig}
\usepackage{natbib}
\usepackage{verbatim}
\usepackage{morefloats} 

\usepackage[caption=false]{subfig}
\usepackage{float}

\usepackage{CJK}

\def\beq{\begin{equation}}
\def\eeq{\end{equation}}
\def\bcm{}

\def\kepler{{\it Kepler}}
\def\keck{HIRES}

\def\het{HRS}

\def\chisq{$\chi_{\nu}^2$}
\def\degree{^{\circ}}

\usepackage{xcolor, fontawesome}
\definecolor{twitterblue}{RGB}{64,153,255}

\begin{document}

\begin{CJK*}{UTF8}{gbsn}

\title{Calibrating Iodine Cells for Precise Radial Velocities}

\correspondingauthor{Sharon Xuesong Wang}
\email{sharonw@carnegiescience.edu}

\author[0000-0002-6937-9034]{Sharon Xuesong Wang (王雪凇)}
\affiliation{Carnegie Observatories, 813 Santa Barbara St., Pasadena, CA 91101, USA}

\author[0000-0001-6160-5888]{Jason T. Wright}
\affiliation{Center for Exoplanets and Habitable Worlds, Department of Astronomy \& Astrophysics, The Pennsylvania State University, 525 Davey Laboratory, University Park, PA 16802, USA}

\author{Phillip MacQueen} 
\affiliation{McDonald Observatory and Department of Astronomy, The University of Texas at Austin, Austin, TX, USA}

\author[0000-0001-9662-3496]{William D. Cochran}
\affiliation{McDonald Observatory and Department of Astronomy, The University of Texas at Austin, Austin, TX, USA}



\author{David R. Doss} 
\affiliation{McDonald Observatory and Department of Astronomy, The University of Texas at Austin, Austin, TX, USA}

\author{Coyne A. Gibson} 
\affiliation{McDonald Observatory and Department of Astronomy, The University of Texas at Austin, Austin, TX, USA}

\author{Joseph R. Schmitt}
\affiliation{Department of Astronomy, Yale University, New Haven, CT 06511 USA}



\begin{abstract}

High fidelity iodine spectra provide the wavelength and instrument calibration needed to extract precise radial velocities (RVs) from stellar spectral observations taken through iodine cells. Such iodine spectra are usually taken by a Fourier Transform Spectrometer (FTS). In this work, we investigated the reason behind the discrepancy between two FTS spectra of the iodine cell used for precise RV work with the High Resolution Spectrograph (HRS) at the Hobby-Eberly Telescope. We concluded that the discrepancy between the two HRS FTS spectra was due to temperature changes of the iodine cell. Our work demonstrated that the ultra-high resolution spectra taken by the TS12 arm of the Tull Spectrograph One at McDonald Observatory are of similar quality to the FTS spectra and thus can be used to validate the FTS spectra. Using the software IodineSpec5, which computes the iodine absorption lines at different temperatures, we concluded that the HET/HRS cell was most likely not at its nominal operating temperature of 70{$\degree$}C during its FTS scan at NIST or at the TS12 measurement. We found that extremely high resolution echelle spectra (R$>$200,000) can validate and diagnose deficiencies in FTS spectra. We also recommend best practices for temperature control and nightly calibration of iodine cells.


\end{abstract}




\section{Introduction} \label{sec:intro}

Precise Doppler spectroscopy with iodine cells as calibrators is an effective way to achieve 1--3~m/s radial velocity (RV) precision \citep{butler1996,butler2017}. As starlight passes through the iodine cell before entering the spectrograph, the iodine inside the gas cell imprints dense and narrow absorption lines on top of the stellar spectrum, providing wavelength and spectrograph calibration and enabling precise RV extraction from the observed stellar plus iodine spectrum. The iodine method has played an important role in exoplanet discovery, such as the detection of the first stars with multiple planets \citep{butler1999}, the first Earth-density planet \citep{howard2013,pepe2013}, and also the characterization of first sample of transiting sub-Neptune and super-Earth planets \citep{marcy2014} which enabled the first studies on the demographics of small exoplanets \citep[e.g.,][]{wu2013, weiss2014, rogers2015,wolfgang2015a,wolfgang2015b}. 

As of early 2019, most of the spectrographs with precise RV capabilities on 8--10 meter class telescopes use iodine cells as calibrators. They are Keck's High Resolution Echelle Spectrometer (HIRES; \citealt{hires1994} and the Hobby-Eberly Telescope (HET)'s High Resolution Spectrograph (HRS; \citealt{1998SPIE.3355..387T}, undergoing instrument upgrade as of early 2019) in the Northern Hemisphere. In the South, there is the Ultraviolet and Visual Echelle Spectrograph (UVES; \citealt{uves2000} and the Planet Finder Spectrograph (PFS) on Magellan \citep{pfs2010}.\footnote{Other spectrographs with precise RV capabilities on 8--10-meter class telescopes are the Echelle SPectrograph for Rocky Exoplanets and Stable Spectroscopic Observations (ESPRESSO; \citealt{espresso2010}) on VLT, which is an optical spectrograph, and three infrared spectrographs --- the Habitable-zone Planet Fidner (HPF) on HET \citep{hpf2012,hpf2019} and the InfraRed Doppler spectrometer (IRD) on Subru \citep{ird2014,kuzuhara2018}} All of these spectrometers have achieved an RV precision of 1--2~m/s, except for HET/HRS, which has an RV precision of 3-5 m/s \citep{2009MNRAS.393..969B}. Our work is motivated by this apparent under-performance of HRS in comparison with other iodine calibrated RV spectrographs such as HIRES.

HET's HRS has multiple settings to meet a variety of science goals, with its RV mode typically having a spectral resolution of $R=60,000$ (and sometimes 120,000). The spectral image is captured by 2 CCDs, covering a spectral range of 4200-11000\AA. Unlike Keck's HIRES, HRS is a fiber fed spectrograph, with choices of science and calibration fibers feeding into the spectrograph and slits with various widths placed at the fiber exit to provide a variety of spectral resolutions. The first planet discovered by \het\ is HD 37605$b$ \citep{cochran2004, wang2012}, and since then it has contributed to several detections of exoplanet systems (e.g., \citealt{2007ApJ...665.1407C,2009ApJS..182...97W,2016A&A...585A..73N}) and performed \kepler\ follow-up (e.g., \citealt{2014ApJ...795..151E}). As fibers provides more stability in the spectrograph response function, or the instrumental profile (IP), in principle, HRS should achieve a higher RV precision than the slit-fed spectrographs such as HIRES.

We began our investigation in the HRS precision problem by examining the fidelity of the calibrator that holds the key to its precise RV capability -- the iodine cell and its atlas, which is the focus of this paper. The iodine atlas, or the iodine reference spectrum, originates from a Fourier Transform Spectrometer (FTS) scan of the iodine cell illuminated by a continuum source (see, e.g., \citealt{crause2018}). It typically has a very high signal-to-noise ratio (SNR) and an extremely high spectral resolution (typically 200,000--500,000). Therefore, the iodine atlas is generally regarded as the ``ground truth" for the cell, which is critical for forward modeling the lower-resolution (60,000) RV spectra \citep{butler1996}. An accurate knowledge of the iodine absorption lines is the basis for accurate and precise calibration for the wavelengths and IP variation, and both are critical for achieving high RV precision from the observed spectra.

The work in this paper is motivated by the fact that we could not model the iodine spectra taken by HRS well enough (meaning consistent within photon-limited noise) using its iodine atlases. There are several potential reasons behind this mismatch: an inappropriate model for the instrumental profile (i.e., the spectrograph response function), modal noise in the fiber, errors in the raw reduction, a genuine mismatch between the true spectrum of the iodine cell and the measured atlases, and so on. Here we investigate the fidelity of the iodine atlases by comparing them with each other and also through a new method using ultra-high resolution echelle spectra and a code that computes theoretically the absorption lines of iodine.

This paper describes our work and findings in diagnosing and validating the FTS atlases of the HRS iodine cell. In Section~\ref{sec:twofts}, we describe the two FTS spectra of the HRS cell and how they differ, which motivated our work. Section~\ref{sec:tull} describes a new way of validating the iodine atlas, using the Tull spectrograph \citep{tull1995} on the 2.7-meter Harlan J.\ Smith telescope at McDonald Observatory. In Section~\ref{sec:fittemp}, we used the software IodineSpec5 \citep{iodinespec5} to determine the temperatures of the iodine cell and to test the hypothesis that temperature variation was why multiple HRS iodine atlases do not agree. Section~\ref{sec:causetemp} explores possibles reasons behind the temperature variation of the HRS cell, and Section~\ref{summary} summarizes the paper and concludes with some recommendations for using iodine cells for precise RV purposes.

\section{Two Different Atlases for the HRS Iodine Cell}\label{sec:twofts}

The first atlas of the HRS iodine cell is from an FTS spectrum taken at the National Solar Observatory at KPNO using the McMath-Pierce 1-meter FTS (nicknamed ``Barbar" for its large size; this FTS has been decommissioned) in 1993 (we refer to this spectrum as the ``KPNO spectrum" here and after).\footnote{Most of the McMath FTS spectra are publicly available at the NSO archive through this FTP: \url{ftp://nispdata.nso.edu/FTS_cdrom/}.} In 2011, we decided to take another FTS spectrum of the iodine cell for the following reasons: 

(1) In our effort to bring HRS to a higher RV precision, we were unable to model the iodine calibration frames to a photon-limited precision, as we could with HIRES spectra. Iodine calibration frames are iodine spectra taken with a continuum lamp or with a hot, fast rotating star with A or B spectral type, and our forward modeling code model them with the iodine atlas as the input or reference spectrum, and then convolve it with an IP model designed for the spectrograph, with free parameters being the IP parameters and parameters for the wavelength solution for the CCD grid (see, e.g., \citealt{valenti1995}). 

(2) the KPNO spectrum was taken almost two decades ago, and during this time the cell may have gone through changes (such as a temperature change, leaking or condensation, etc., though unlikely, since the cell was designed to be stable).

Therefore, we arranged the HET/HRS cell to be sent to the Atomic Spectroscopy Group at the National Institute of Standards and Technology (NIST) facility in Gaithersburg, Maryland, USA, and obtained a new FTS spectrum on November 15, 2011.\footnote{The serial numbers for these NIST FTS scans are: I111511.005 (a sample spectrum), I111511.006 (the cell at $70\degree$C), I111511.007 (the cell at $75\degree$C), I111511.008 (the cell at $65\degree$C).} A close comparison between this new spectrum from NIST and the old spectrum from KPNO reveals that they
have many differences:
\begin{itemize}
  \item The overall line depths are very different --- the NIST spectrum
    has deeper lines in general.
  \item The absolute wavelength solutions are different, and the
    drifting of wavelength solution or the dispersion scales at
    different wavelengths are also different.\footnote{The wavelength difference is to be expected, as even spectra from the same FTS instrument could differ in wavelengths as the instrument could have small changes over time. Please see \cite{nave2011} for more on wavelength calibration of FTS.}
  \item Even after we adjust the normalization or continuum level of the NIST
    spectrum (assuming the FTS data has normalization issues, contamination from a continuum source, or low
    frequency noise/offset), the line ratios of the two FTS spectra still
    exhibits differences.
\end{itemize}

\begin{figure}[!th]
\centering
\includegraphics[angle=0.,scale=0.5]{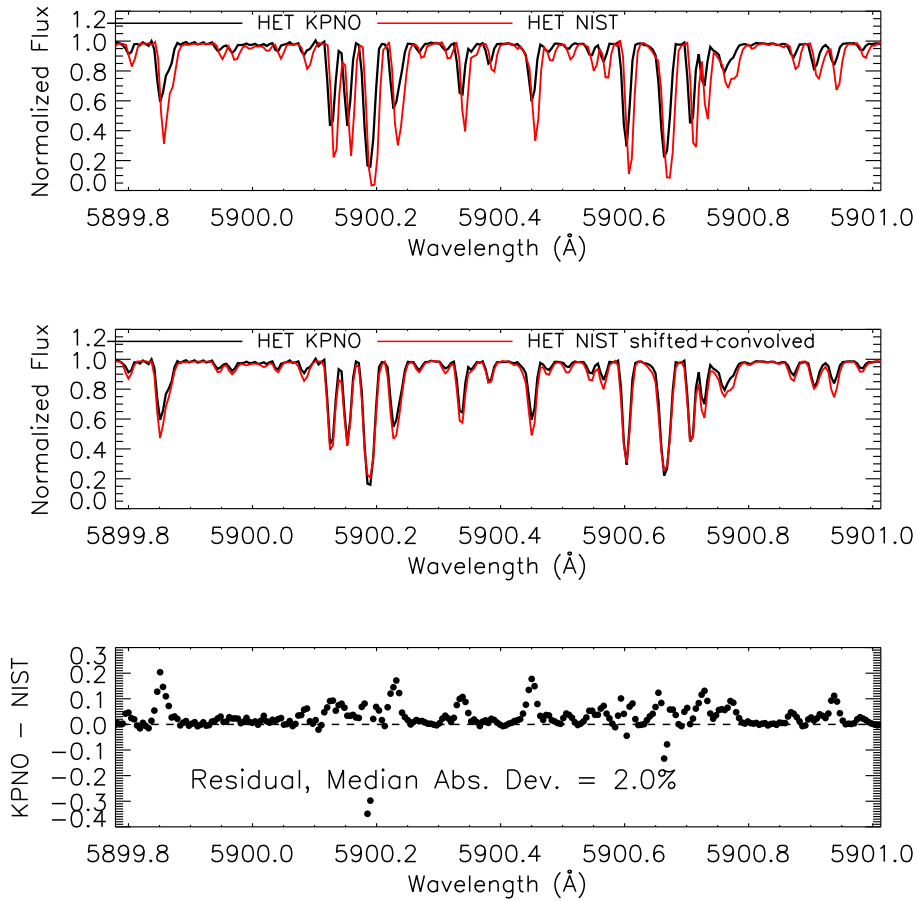}
\caption{Comparison of the KPNO FTS spectrum (black) and the NIST FTS spectrum
  (red) for the HRS iodine cell for a selected
  1.5\AA\ chunk. \textbf{Top:} Two FTS spectra at their original resolution
  and original wavelength solution. \textbf{Middle:} Comparison of the
  two FTS spectra after adjusting the normalization, shifting the wavelengths, and
  applying convolution for the NIST spectrum to match the KPNO spectrum for a more
  direct comparison of line depths/ratios. \textbf{Bottom:} Residuals
  of the middle panel, i.e., the adjusted NIST spectrum minus the KPNO spectrum. The
  median absolute deviation between the two spectra is 0.02
  (2\%), though at many places, especially at line centers, the two
  can differ by up to 20\%.
  \label{fig:fts_old_new}}
\end{figure}

Figure~\ref{fig:fts_old_new} shows the comparison between the two
FTS spectra in a selected 1.5\AA\ region. As the two FTS spectra also differ in
resolution (the NIST spectrum has a higher resolution), the middle panel
is a more direct comparison: the NIST spectrum has been convolved down to
the same resolution with the KPNO spectrum; it is also shifted in
wavelength space so that the two FTS spectra match in absolute wavelength
solution; and it is adjusted to a different normalization level to
match with the KPNO spectrum as much as possible in order to compare their
relative line ratios.

\begin{figure*}[!th]
\includegraphics[angle=0.,scale=0.33]{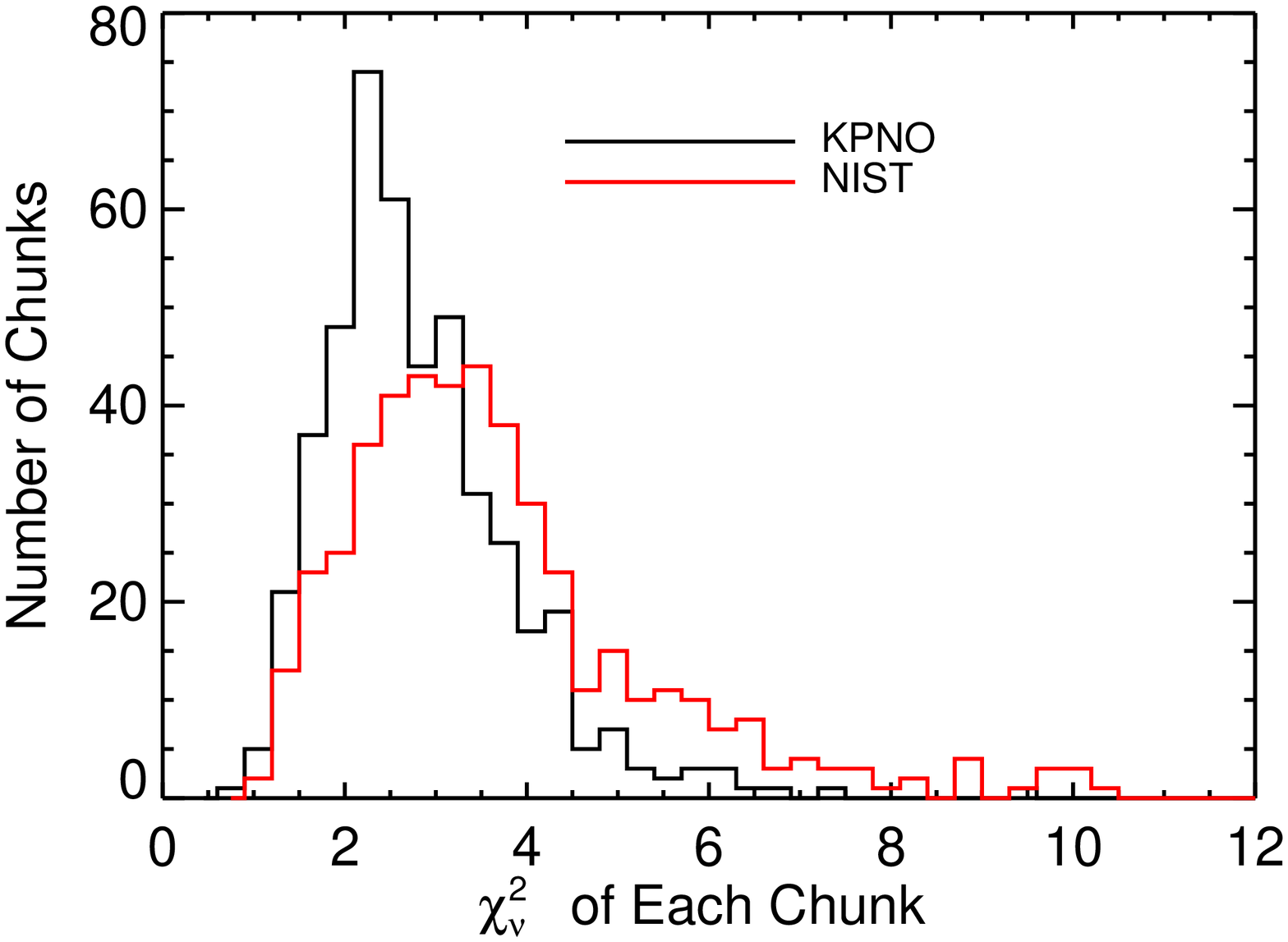}
\includegraphics[angle=0.,scale=0.33]{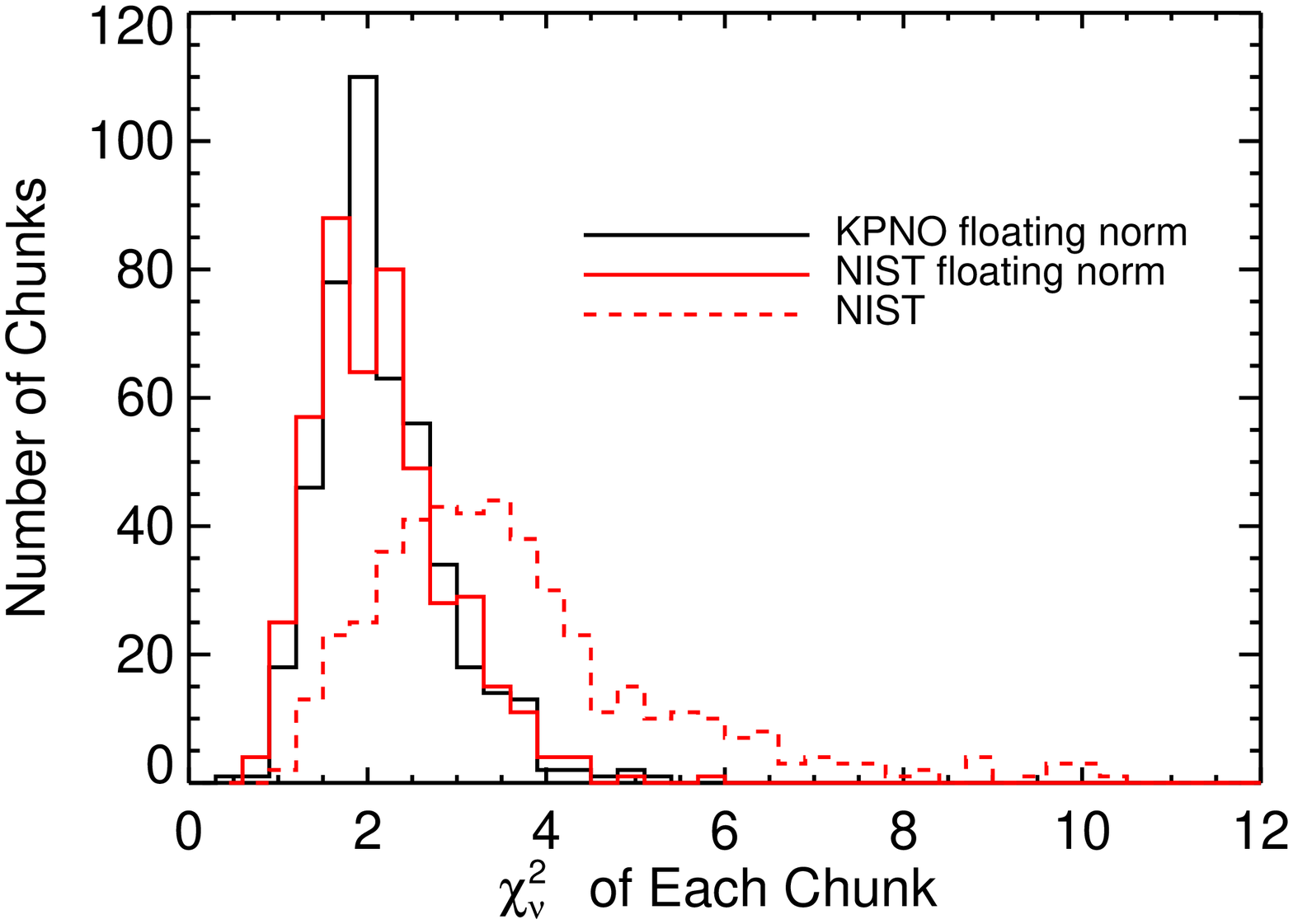}
\caption{Both plots are histograms of \chisq\ values of all the spectral chunks in a single
  iodine calibration frame taken by \het. Each \chisq\ value in the histogram represents
  the \chisq\ goodness of fit for a $\sim$2\AA\ spectral chunk in this selected
  iodine spectrum (each iodine spectrum is divided into several
  hundred of chunks and is fitted independently using a Levenberg-Marquardt least $\chi^2$ fitter).
  \textbf{Left:} \chisq\ histograms for the fit of the iodine
  observation using the KPNO spectrum (black) or the NIST (red) spectrum as the model input iodine
  reference spectrum, respectively. The KPNO spectrum performs better.
  \textbf{Right:} \chisq\ histograms for the two FTS spectra, but both with
  spectral normalization offset for the atlas as a free parameter for each chunk. The two spectra now
  perform at essentially the same level. Dashed red line is the same
  red histogram as plotted in the left panel. Notably, the KPNO spectrum
  also performs better when we float the normalization offset.
  \label{fig:chisq_old_new}}
\end{figure*}

We initially suspected that the NIST spectrum was problematic. The reason is
illustrated in the left panel of Figure~\ref{fig:chisq_old_new}, where
it shows the histogram of \chisq\ values for fitting a selected iodine
calibration frame (a $R=60,000$ lamp observation through the iodine cell) using the two FTS spectra, respectively. Each \chisq\ value is
for a 2\AA\ chunk in this selected iodine spectrum. It is clear
that the NIST spectrum provides a worse fit.

Since the middle panel of Figure~\ref{fig:fts_old_new} presents a much better match between the NIST spectrum and the KPNO spectrum, we decided to add a free parameter to account for an offset term in the spectral continuum of the iodine atlas, in case any of the two FTS spectra has problems in continuum normalization from data reduction or has contamination in the continuum, for example.\footnote{Problems in continuum normalization in the FTS spectra could be due to, for example, the lack of background subtraction. This should not be a problem since the background spectra in the FTS spectra should be negligible. However, this possibility cannot be completely ruled out because background spectra are almost never taken with FTS scans of iodine cells (Dr.~Gillian Nave, NIST, private communications; because they should be negligible and also, background spectrum measurement is just as time consuming as a FTS scan for an iodine cell and thus unpractical).} The right panel of Figure~\ref{fig:chisq_old_new} shows the \chisq\ histograms for the same iodine observation using the two FTS spectra, but adding a free parameter as the normalization offset when fitting each chunk (note: this normalization offset parameter is a free parameter for each chunk, not a single global parameter). The two FTS spectra then performed at essentially the same level in term of goodness of fit for an iodine frame, with the KPNO spectrum performing slightly better.

This was both encouraging and worrisome at the same time. It was encouraging because it seemed that we have found the problem with the NIST spectrum (i.e., the continuum) and also had a solution for it. It was very worrisome because this revealed that:
\begin{itemize}
  \item Even the KPNO spectrum performed visibly better when we floated the
    normalization parameter. This may suggest that there are
    normalization issues or low frequency errors/noise in the KPNO
    spectrum as well.
  \item Obtaining high-quality, reliable FTS spectra of iodine cell is
    not an easy task, and the FTS spectra cannot be naively trusted as the
    ``ground truth'', ultra-accurate templates of the complicated iodine
    spectrum.
  \item The reason why adding a floating normalization fitted the
    data better might be because it accounted for optical depths difference
    between the atlas and the actual observations, which may be a result
    of changes in the cell temperature or the iodine column density.
  \item The forward modeling code, when floating normalization as a free parameter,
    could not distinguish which FTS spectrum was more accurate (by comparing \chisq)
    even when the two FTS spectra differed as much as $\sim$5--10\% at places and
    also had obvious line ratio differences (see comparison in bottom
    panel of Figure~\ref{fig:fts_old_new}). However, this level of
    difference in FTS may affect the RV precision, and not knowing
    which atlas is the correct one definitely affects our ability to
    search for a better IP model (which was also suspected to be a limiting factor for HRS; see \citealt{thesis}) and improve the RV precision of \het.
\end{itemize}  

\begin{figure}[!th]
\centering
\includegraphics[angle=0.,scale=0.38]{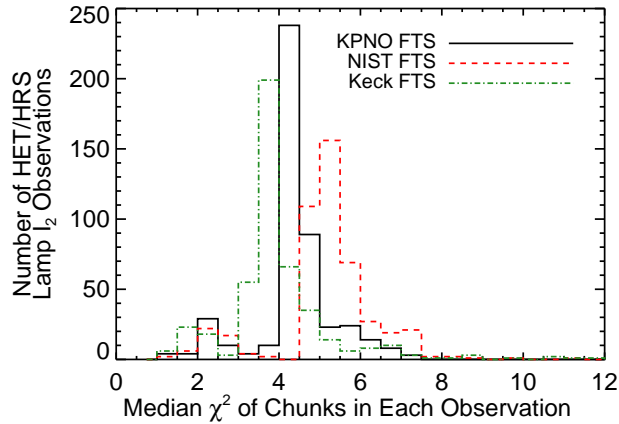}
\caption{Comparison of the median \chisq\ values for the fits of HRS iodine
  frames using the \het\ KPNO spectrum (black solid line), the
  HRS NIST spectrum (red dashed), and the {\bf HIRES} KPNO spectrum (green
  dotted-dashed). Each data point represents the median of the \chisq\ values
  of all the chunks in a single iodine spectrum (these are all
  lamp-illuminated iodine calibration frames, not from A/B stars, i.e., through the HRS calibration fiber instead of the science fiber). Over 500 iodine calibration frames were used for this plot to ensure the statistical significance. The \keck\ KPNO spectrum provided a better fit than the
  both \het\ FTS spectra when fitting \het\ iodine observations.
  \label{fig:lampi2fit}}
\end{figure}

It was perhaps even more alarming and more puzzling that, when we used the KPNO
spectrum for the {\bf HIRES iodine cell} to fit an {\bf HRS} iodine
frame, it yielded smaller \chisq\ values than using any of the
other two FTS spectra of the HRS cell (Figure~\ref{fig:lampi2fit}). The \het\ cell's KPNO spectrum
was taken at the same time using the same FTS machine as the spectrum for the \keck\
cell \citep{butler1996}. However, the temperatures of these two cells were very
different: the \keck\ cell was designed to work at 50~$\degree$C, while
the \het\ cell was designed to work at $70\degree$C. A closer look
revealed that the HRS KPNO spectrum and the HIRES KPNO spectrum had very
similar line depths, with the HIRES cell's FTS spectrum having slightly deeper lines (due to a
higher iodine molecule column density; see more later in Section~\ref{sec:fittemp}).

The findings above prompted us to seek an independent way to perform quality checks for any FTS spectrum --- not just comparing their relative qualities or performances. One natural choice would be to obtain spectra taken with high-resolution echelle spectrographs, which are measurements of the iodine spectrum directly in the real wavelength space instead of in the Fourier space, and thus they could serve as good reference spectra as they would suffer from different types of error compared with FTS spectra. Since FTS spectra are usually at a very high spectral resolution ($200,000$--$500,000$), this limited our choice to essentially only one spectrograph at the time (circa 2014) --- the TS12 configuration of the Tull Spectrograph One (TS1) at the 2.7-meter Harlan J.\ Smith Telescope at McDonald Observatory \citep{tull1995}.

Our goal for taking spectra with TS12 was to answer the following questions: {\bf which FTS spectrum better describes the \het\ iodine cell: the KPNO one, or the NIST one? And why? Why did the FTS spectrum for the HIRES cell work even better than the HRS FTS spectra?} While adding an additional normalization offset parameter provided better fits for the iodine frames, we could not afford to add such an additional parameter when extracting RVs from star plus iodine spectral data -- this normalization offset parameter would be degenerate with the Doppler shift and the wavelength solution parameters and thus would lead to a decrease in the RV precision. Therefore, we needed to resolve the issue instead of adding an additional free parameter to mend the FTS spectrum.

\section{Validating FTS spectra of Iodine Cells using the Tull Spectrograph One}\label{sec:tull}

To break the tie between various FTS spectra, we used the TS12 configuration of the Tull Spectrograph One \citep{tull1972}. The Tull spectrograph with TS12 was the highest resolution spectrograph on sky at the time of our experiment around 2013. It employs a double pass on the echelle grating to achieve a resolution of $\sim$500,000, based on ThAr line measurements taken during our observations. TS12 was frequently used for studying lines of the interstellar medium in the 1990s \citep[e.g.,][]{sembach1996}, and was rarely used in the recent years.

We had two rounds of TS12 observing session, both during day time and when the telescope was scheduled for Cassegrain instruments and the Tull Spectrograph in the coud\'e room was free for use. For the first run (from September 7 to September 9, 2013; done by Sharon X.\ Wang and Ming Zhao), we measured the iodine absorption spectrum for the iodine cell of the Sandiford Spectrograph at the 2.1-meter telescope at McDonald Observatory, because the HRS cell was unavailable (it was still under active use for planet search programs at HET). The main purpose of the first run was to validate the quality of the TS12 spectrum and whether we could use it to validate the FTS spectra. The Sandiford cell also had a KPNO FTS spectrum, which was taken together with the KPNO spectrum of the HRS cell in 1993, so it also served the purpose of testing the overall quality of the KPNO FTS spectra.

In the second TS12 run, we measured the spectrum for the \het\ iodine cell in its original enclosure and temperature controller (the enclosure and the thermal control system have been rebuilt since then). We also took spectra of other iodine cells, including the MINERVA \citep{minerva2015} iodine cell and the iodine cell used at the McDonald 2.7-meter Smith telescope. The run was from October 13 through October 16, 2014 and was carried out by Ming Zhao, Kimberly M.~S.~Cartier, and Joseph R.~Schmitt, with the help of Phillip MacQueen in setting up the cells.

The hardware settings and data reduction methods are the same for both
runs, which are described below.

\textbf{\textit{Hardware Settings:}} We used the TS12 configuration of the Tull
Spectrograph One, and the specific instrument settings we used are listed in
Table~\ref{tab:hardware}. Slit \#23 is chosen to maximize SNR while
maintaining sufficient resolution --- it is among the longest slits and
is the second narrowest slit. The slit is $0\arcsec.32$ wide and $30\arcsec$ tall, or 5.8 by 543 pixels, given the plate scale of TS12 with the TK4 CCD being $0.55\arcsec$ per pixel. The Sandiford cell was kept at a
temperature of 49.9--50.1$^\circ$C, the same as its working
temperature for RV observations and also the same as its temperature when the KPNO was taken
(50~$\degree$C). The \het\ cell was measured at four different
temperatures: room temperature, 50~$\degree$C, 60~$\degree$C, and
70~$\degree$C (its working temperature). 

\renewcommand{\arraystretch}{1.3} 
\begin{deluxetable}{rl}
\tablecaption{Hardware Settings for TS12\label{tab:hardware}}
\tabletypesize{\scriptsize}
\tablehead{\multicolumn{2}{c}{Tull Spectrograph One, TS12, Coude107} }
\startdata
  Echelle & E1 \\
  Cross Disperser & c \\
  CCD & TK4, 1024$\times$1056 \\
  On-chip Binning & 1$\times$1 \\
  Slit & \#23 (L$\times$W $=30\arcsec \times 0\arcsec .32$) \\
\enddata
\end{deluxetable}

\begin{figure}
\centering
\includegraphics[scale=0.35]{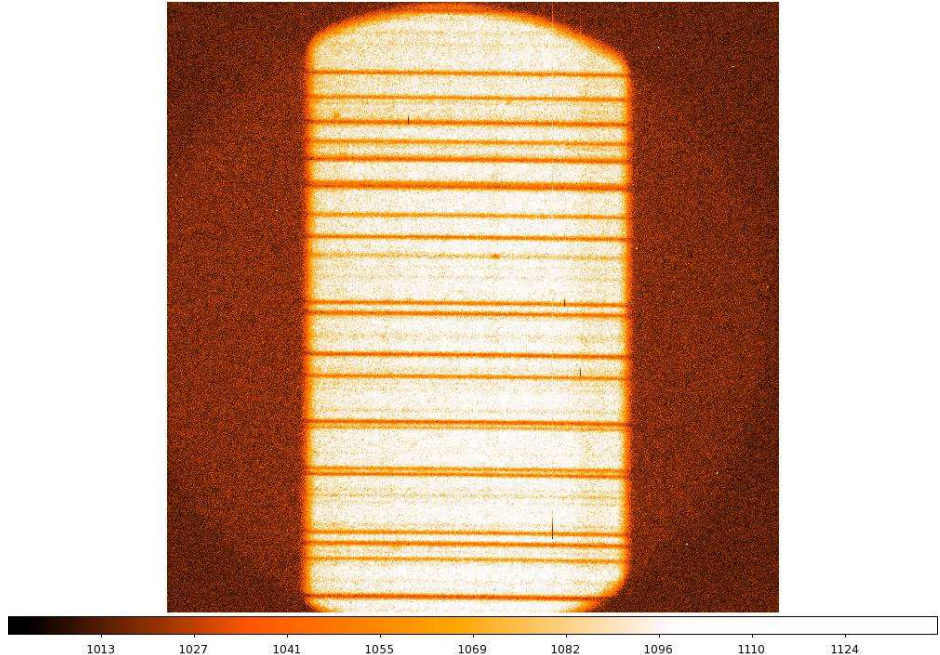}
\caption{One raw image frame taken using the TS12 setting of Tull
  Spectrograph One. It contains about 1.9\AA\ of iodine absorption spectrum. 
\label{het:fig:ts12image}}
\end{figure}

\textbf{\textit{Observations:}} A single exposure frame for the iodine
spectrum covers about 1.9\AA\ (Figure~\ref{het:fig:ts12image}). The
dispersion direction runs vertically along the chip with increasing
wavelength when increasing the $y$-axis pixel. The dispersion scale is
about 0.002\AA\ per pixel, with $\sim$7 pixels per resolution element. We
immediately preceded or followed each exposure with a flat fielding
frame in the same echelle and cross disperser positions as the iodine frame. 
The exposure times for the iodine and flat frames are both 45
seconds (90 seconds for the \het\ cell) to achieve a signal-to-noise ratio
(SNR) of 160 per CCD pixel (again, higher for the \het\ cell). Neighboring frames
differ by about 1\AA\ in absolute wavelength, leaving ample overlapping regions 
between frames for post-reduction stitching of the spectrum. If prominent Solar or
ThAr line was predicted within the wavelength coverage of a frame,
then we also took a Solar or ThAr frame to verify the rough wavelength
solution. The exposure times for these Solar or ThAr frames varied --- typically 
from a couple minutes to up to 10 minutes. We took dark frames (45s each, about 10 frames) in
the morning at the beginning of each day.

\textbf{\textit{Reduction:}} We combined and averaged all available
dark frames and created a master dark frame. Then we subtracted the
master dark from all flat and iodine frames. After outlier rejection
(cosmic rays, chip defects, etc.), we modeled the scattered light for
each row of pixels by using the region on the CCD outside of the slit image. To estimate the scattered light in an image row, we added the flux in 160 neighboring rows (80 above and 80 below the target row) along the column direction (i.e., the dispersion direction). We then fitted a third order polynomial to the added flux outside of the slit image region to construct a model of the scattered light. Then we evaluated the values of scattered light within the slit image region using the best-fit polynomial and subtracted it off. Both the flat and iodine frames have scattered light removed. Then we normalized the flat frames and divided each iodine frame by its associated normalized flat (for the slit image regions only).

\textbf{\textit{Extraction:}} As the slit does not lie perfectly along
the $x$-axis direction on the chip, we rectified the image before extraction by first taking
columns along the dispersion direction and cross-correlating these
columns. The relative shift between columns was determined via oversampling the two spectra by a factor of 10, and then performing cross correlation to determine the amount of shift by fitting a Moffat function to the cross correlation function. Then we interpolated the columns via spline interpolation to a common pixel grid after shifting them to create an aligned image. We then added the flux along the row direction (i.e., cross-dispersion direction) and obtained the reduced, extracted spectrum. Each spectrum is then
normalized by dividing the estimated continuum (top 5\% counts). Due
to lower quality of scattered light removal near the edge of the chip,
we discarded the top 80 and bottom 80 rows of pixels. Thus the
extracted spectrum from each frame is about 1.6\AA\ across (instead of
1.9\AA). The reduced frames are then stitched together by finding
the overlapping region through cross correlation of each pair of
neighboring frames while taking into account the changes and differences
of dispersion scales across frames.

\textbf{\textit{Mapping onto FTS:}} The extracted TS12 spectra are not wavelength calibrated naturally like the FTS spectra. To compare with the FTS
spectra, we divided the TS12 spectrum into 2\AA\ chunks and projected
each chunk onto the FTS spectrum via cross correlation. In this way we
obtained the absolute wavelength solution and dispersion scale (as set
by the wavelength solution of the FTS spectrum) for the TS12
spectrum.

\begin{figure*}[!th]
\includegraphics[angle=0.,scale=0.33]{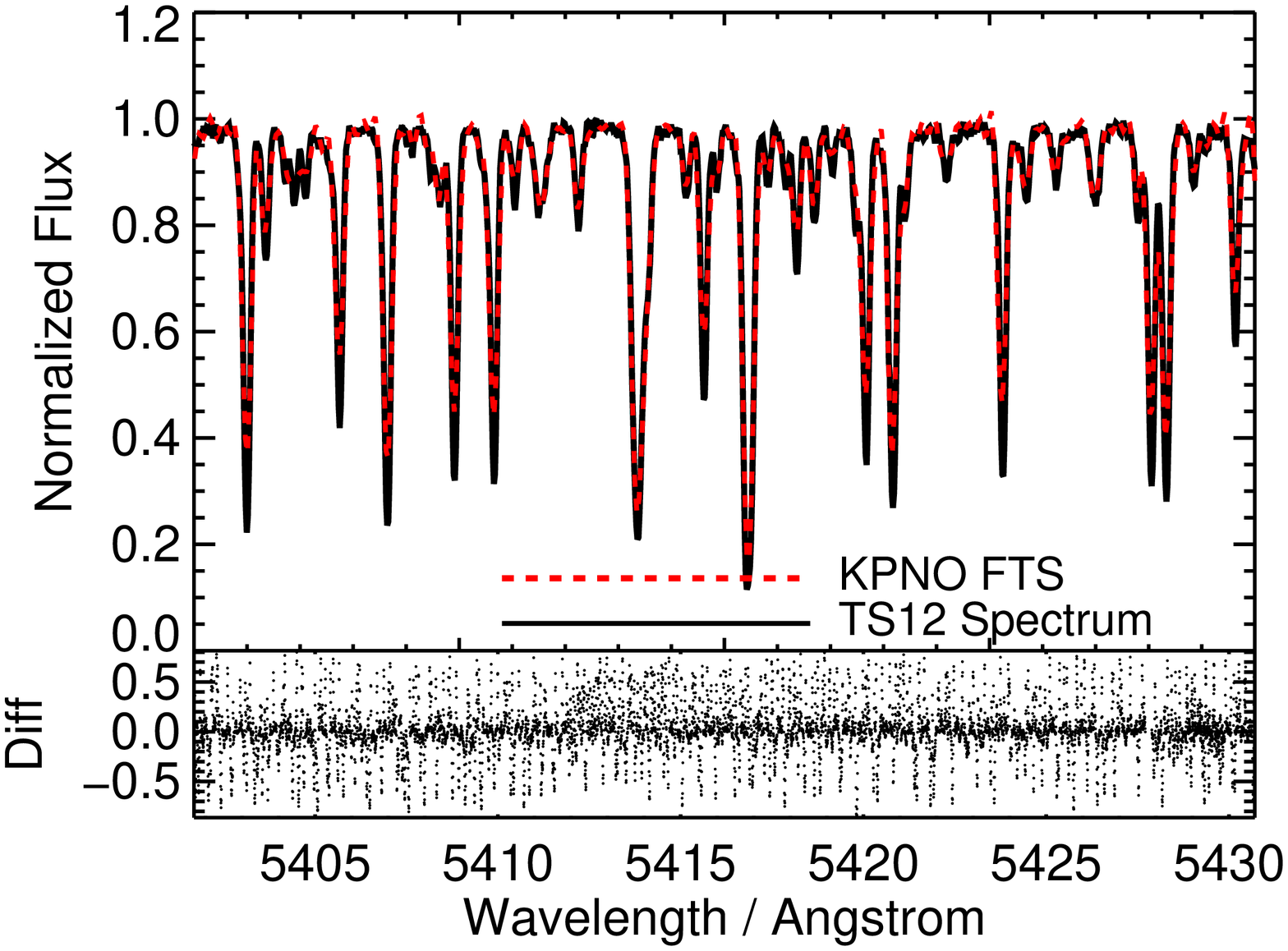}
\includegraphics[angle=0.,scale=0.33]{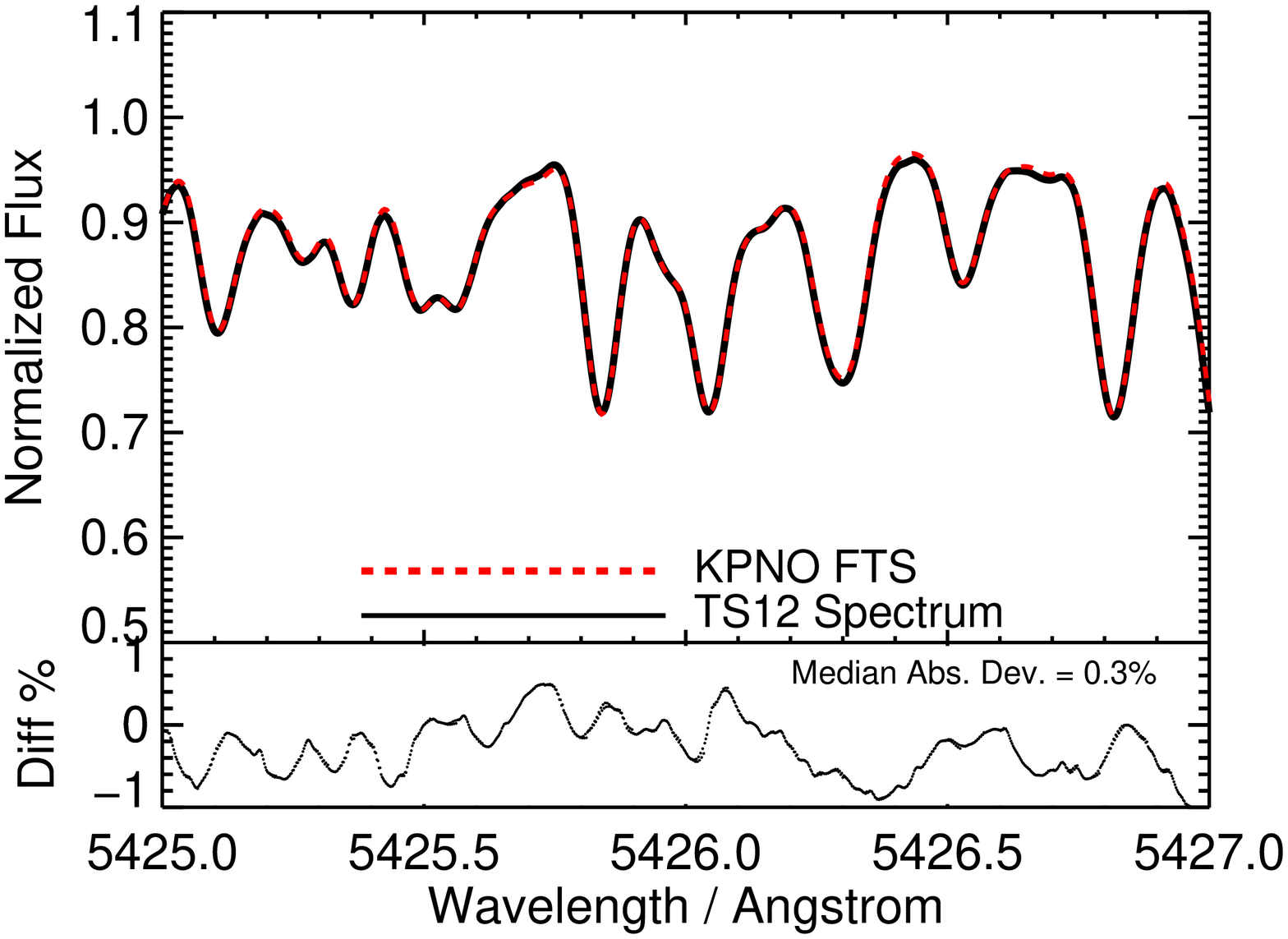}
\caption{Comparison of the Sandiford iodine cell KPNO FTS spectrum and
  the spectrum taken with TS12. \textbf{Left:} Comparison of the two
  spectra in their native resolutions (both about
  $400,000$--$500,000$). \textbf{Right:} Comparison of the two spectra
  convolved down to about $60,000$ resolution, which is the resolution
  of typical iodine observations or radial velocity observations
  (star$+$iodine). Bottom panel shows the residuals in percentage of
  the TS12 spectrum minus the KPNO spectrum, with a median absolute
  deviation of 0.3\%.
  \label{fig:ts12}}
\end{figure*}

The results from our first TS12 run using the Sandiford cell
demonstrated that an iodine cell spectrum taken with TS12 has the same
quality as an FTS spectrum to serve as the ``true solution" of the iodine
spectrum in the forward modeling process for RV extraction (although a wavelength 
solution would have to be derived). The left panel of Figure~\ref{fig:ts12} shows a direct
comparison of the reduced TS12 spectrum (a random 2\AA\ chunk) with
the KPNO FTS spectrum, at their native resolutions.\footnote{Note that the
TS12 spectrum appears to have a higher resolution than the FTS
spectrum. According to the header of the FTS spectrum, its resolution is about
$491,000$. An FFT analysis on the TS12 spectrum (to see where the
high-frequency signal cuts off and becomes indistinguishable from the
noise) shows that its resolution is about $455,000$ and maybe even
higher.}

\begin{figure}[!th]
\centering
\includegraphics[angle=0.,scale=0.45]{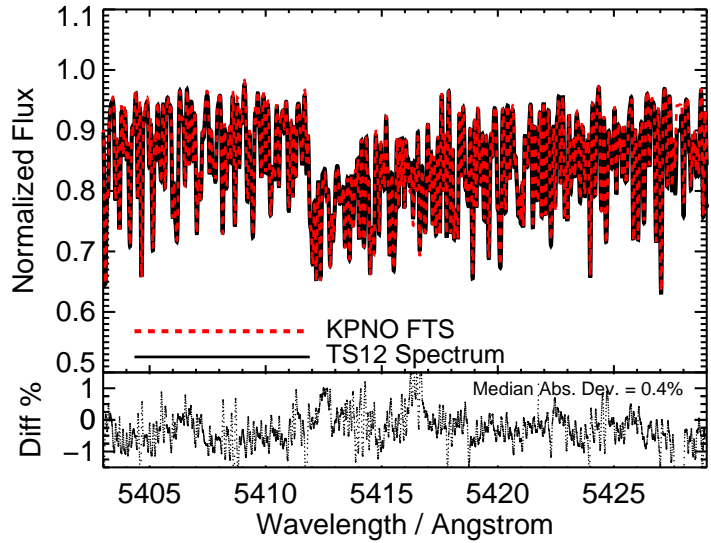}
\caption{The same as the right panel of Figure~\ref{fig:ts12}, the
KPNO spectrum and the TS12 spectrum for the Sandiford cell both at
$60,000$, but for the entire $\sim$30\AA\ TS12 spectrum available. 
  \label{fig:60k_all}}
\end{figure}

To make a more direct comparison and also to see the differences of
the two spectra (if any) would make a significant impact when fitting
a typical iodine observation with a spectral resolution of $60,000$, we degraded the resolution
of both spectra to $60,000$ by convolving them with a Gaussian IP. 
The right panel of Figure~\ref{fig:ts12} illustrates the
comparison of the two spectra at $R\sim60,000$, with residuals of the
TS12 spectrum minus the KPNO FTS spectrum plotted in the bottom
panel. The two spectra differ by a median absolute deviation of 0.3\%
($0.4\%$ for the entire $\sim$30\AA\ spectrum available as shown in
Figure~\ref{fig:60k_all}).\footnote{ For comparison: when fitting the
HET/HRS Iodine observation used for creating
Figure~\ref{fig:chisq_old_new} (median SNR for a typical chunk is
$\sim$150 per pixel, or 0.65\% RMS in shot noise), for a typical chunk, the median
absolute deviation between the observation and the best-fit model is
0.73\% (the RMS value is 1\%, thus \chisq\ is $\sim 2$--$3$).  } As
the TS12 spectrum has a SNR of about 160 per pixel and we have convolved the
comparison spectrum down to $R\sim60,000$, the expected shot noise
should be $\sim 1/160\times \sqrt{450,000/60,000}=0.23\%$. The
additional $\sim 0.1\%$--$0.2\%$ of noise may come from flat fielding,
scattered light removal, cosmic ray removal and interpolation between
pixels, stitching of spectra, projection onto the FTS spectrum and
interpolation for comparison purposes, and so on.

\begin{figure}
\centering
\subfloat{\includegraphics[scale=0.32]{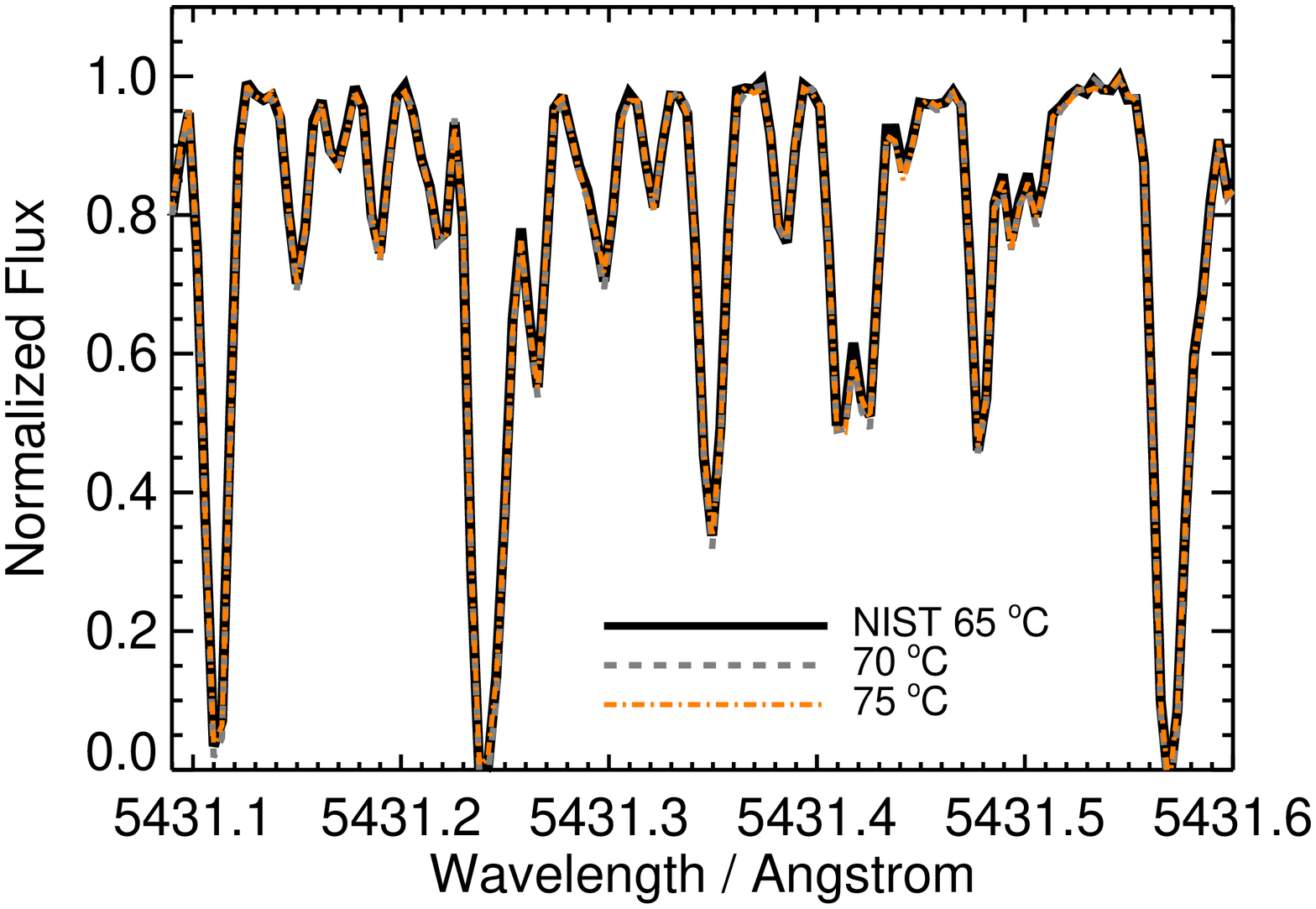}}
\subfloat{\includegraphics[scale=0.32]{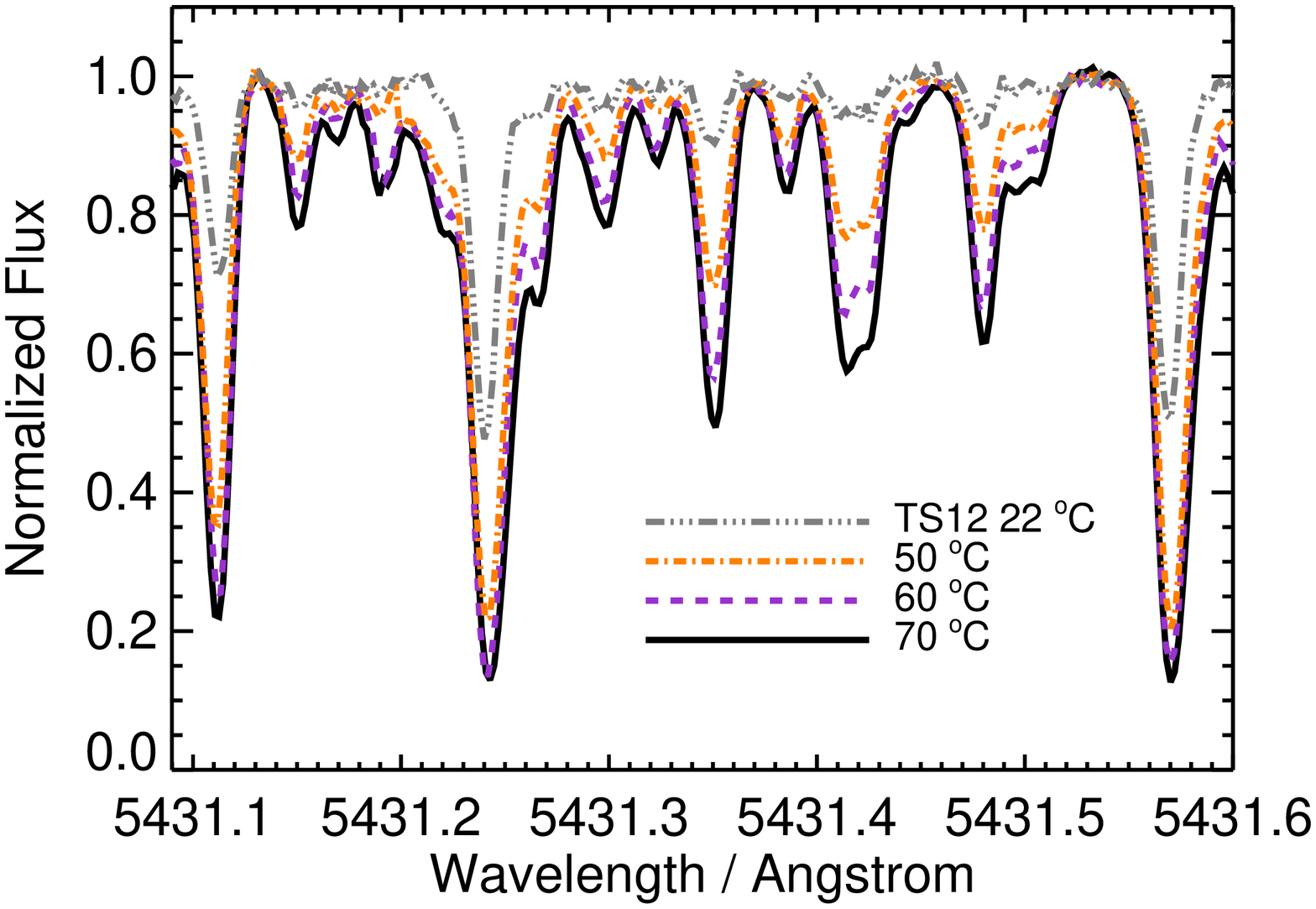}}
\caption{{\bf Left:} \het\ cell NIST FTS spectrum at three different
temperatures. {\bf Right:} \het\ cell TS12 spectra at four different
temperatures for the same wavelength region, which, unlike the NIST
FTS spectra, shows significant difference when the temperature of the cell
changes by 10~$\degree$C.
\label{het:fig:tempchange}}
\end{figure}

After demonstrating that we can use TS12 spectrum to validate FTS
spectra using the Sandiford cell, we performed our second TS12 run with the \het\ cell (and the
2.7-meter cell and the MINERVA cell). For
the 2.7-meter cell, its TS12 spectrum matches very well to its FTS atlas, similar to the results with the Sandiford cell, proving again that TS12
is an appropriate tool for validating FTS atlases. For the HRS
iodine cell, the story is very different and the results are very informative:

(1) Assuming that the \het\ cell temperature control was reliable
during our TS12 run, then temperature change on the order of 5--10~$\degree$C in iodine cell
should induce a visible change in the absorption spectrum (right panel
of Figure~\ref{het:fig:tempchange}). On the other hand, the
temperature of the iodine gas in the cell does not seem to be at $70\degree$C as set 
by the temperature controller during the NIST spectrum (left panel of 
Figure~\ref{het:fig:tempchange}; see more in Section~\ref{sec:fittemp}). This could explain the difference
between \het\ cell's NIST spectrum and the KPNO spectrum. The HRS cell during the NIST FTS scan appears to have stayed at a temperature higher than $70\degree$C during the 
entire FTS scan, at the same high temperature with three different 
temperature settings at 65, 70, and $75\degree$C. The \keck\ cell was
also scanned at three different temperatures (50, 60, and
70~$\degree$C) at KPNO in 1993, and there are also visible differences
between these three sets of FTS spectra, unlike HRS cell's NIST spectrum.

\begin{figure}
\centering
\includegraphics[scale=0.5]{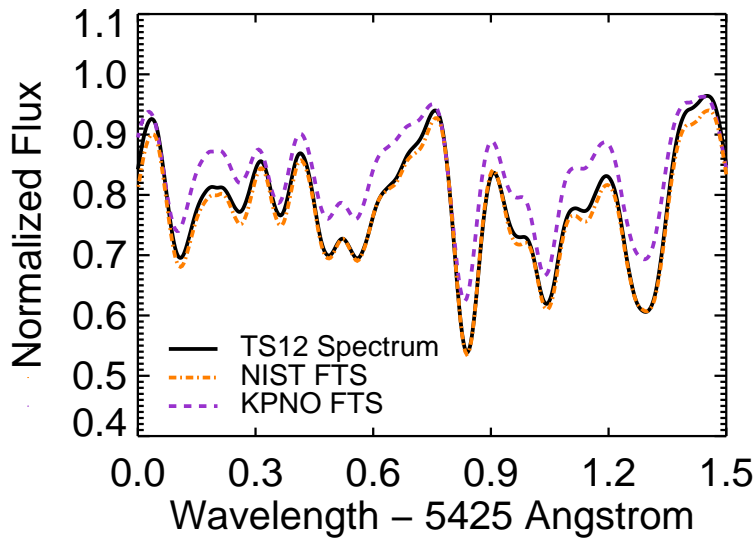}
\caption{TS12 spectrum (black solid line) vs.\ NIST FTS (orange
  dot-dashed) vs.\ KPNO FTS (purple dashed) for the \het\ iodine
  cell at 70~$\degree$C, all convolved down to a resolution of R $=$
  60k (the same as a typical \het\ observation) for comparison
  purposes. The TS12 spectrum matches the NIST FTS better, having
  deeper lines compared to the original KPNO FTS. The remaining
  difference between NIST FTS and the TS12 spectrum might be due to
  differences in cell temperatures or other changes with the cell. 
\label{het:fig:hetts12}}
\end{figure}

(2) The TS12 spectrum at $70\degree$C matches better with the more
recent but potentially problematic NIST FTS spectrum
(Figure~\ref{het:fig:hetts12}). In fact, the TS12 spectrum does not match the NIST FTS spectrum well
enough given the high SNR nature of both spectra, so they are actually inconsistent, although less so compared with the KPNO FTS spectrum. If the HRS cell was indeed at $70\degree$C during the KPNO spectrum, as we have seen that the KPNO FTS spectra are validated using the TS12 spectra for both the Sandiford cell and the 2.7-meter cell, then Figure~\ref{het:fig:hetts12} suggested that the HRS cell was not at its set temperature of $70\degree$C during either the TS12 run or the NIST spectrum. Or, perhaps the amount of iodine vapor was somehow different.

These results prompted us to resolve for a second route to try to break
the tie: using synthetic iodine absorption spectra, which is described
in the next subsection.

\section{Measuring iodine cell temperatures using IodineSpec5\label{sec:fittemp}}

Using the TS12 spectra, we found that the temperature of the iodine
gas in the cell might not be the same one as reported by the temperature
controller. However, we still could not break the tie between the KPNO
spectrum and the NIST spectrum for the HRS cell: the KPNO spectrum provided a better
fit to real observed data, but the TS12 spectrum showed us that the
NIST spectrum looked closer to the ``truth'' as defined by TS12 (if that was the truth). Nor did we
understand why the KPNO spectrum for the \keck\ cell worked the best for
\het\ data.

To answer these questions, we found a second venue that provided
reliable, ultra-high resolution, and wavelength calibrated iodine
atlas -- a theoretical code that computes synthetic iodine
transmission spectrum (at any specified temperature) based on both
quantum physics and empirical calibrations (IodineSpec5;
\citealt{iodinespec5}).\footnote{Our copy of the IodineSpec5 program was kindly provided by \citeauthor{iodinespec5} to the authors in October 2014. It runs on Windows machines.} The direct output of the code contains arrays of wave numbers and, effectively, opacity ($\alpha$) for user-specified
iodine isotope mix (for our purposes we only add $^{127}I_2$),
temperature, wave number range, and a line broadening kernel
(we chose thermal/Gaussian). To use the output of IodineSpec5 to fit
an actual iodine absorption spectrum, there are two parameters:
gas temperature and a constant which scales with iodine molecular
column density, which we simply refer to as the column density hereafter.

\begin{figure}
\centering
\includegraphics[scale=0.5]{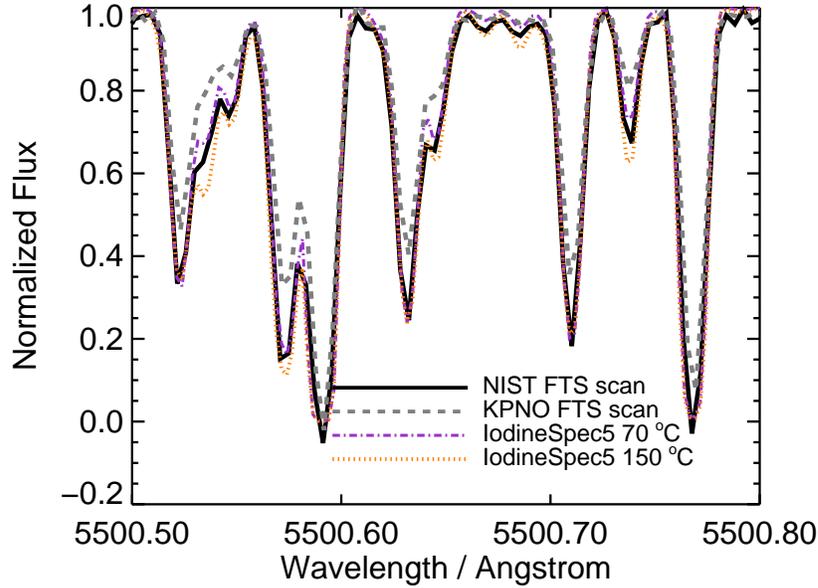}
\caption{NIST FTS (black solid line) and KPNO FTS (gray dashed) spectra 
for the HRS cell compared with theoretically computed iodine lines by IodineSpec5 at
70~$\degree$C (purple dot-dashed) and 150~$\degree$C (orange dotted). All
spectra are at their original resolution. There are two free
parameters for the theoretical lines: temperature and iodine column
density. For this plot, we optimized the iodine column density for the
theoretical lines at both temperatures to try to match the NIST FTS. As
illustrated, neither the 70~$\degree$C nor the 150~$\degree$C spectrum 
can produce a good match, and, as shown later, the
best-fit temperature is around 110~$\degree$C. Note that the
theoretical lines and the NIST spectrum have different broadening
kernels.
\label{het:fig:nisteyeball}}
\end{figure}

A quick comparison between the NIST spectrum and the IodineSpec5 models revealed
that the NIST spectrum seems to be around 110~$\degree$C, based on
visually examining the line ratios
(Figure~\ref{het:fig:nisteyeball}). However, the synthetic iodine
spectrum and the FTS spectra have different broadening kernels. In order
to ``fit'' the NIST spectrum with the synthetic spectra at various
temperatures, we convolved the NIST spectrum with a single Gaussian kernel
with $\sigma=0.0078$ (roughly at $R=$ 200,000 to match the
\keck\ cell KPNO spectrum for comparison and illustration
purposes). Except for the \keck\ cell's FTS spectrum, we did the same to lower the
resolution of other FTS spectra or TS12 spectra when using IodineSpec5
to determine their best-fit temperatures. There are four parameters in our
fit: temperature, column density, resolution ($\sigma$ for the single
Gaussian kernel to convolve with the synthetic IodineSpec5 spectrum), and a
wavelength shift (since it is common for different FTS spectra to have 
offsets in wavelengths due to calibration errors). We first optimized the column density, $\sigma$, and
the wavelength shift while fixing the temperature (using a levenberg-Marquardt
least-$\chi^2$ fitter with the {\tt mpfitfun} package in IDL), and then
we compare the goodness of fit of each model at different temperatures
to determine which temperature best describes the FTS spectrum or the TS12
spectrum at hand. The reason for this two-step optimization is that we had
to generate the IodineSpec5 model spectra on a discrete temperature
grid.

We first fitted the \keck\ cell KPNO FTS spectrum, whose temperature was
known (50~$\degree$C) and probably reliable. We knew that this FTS spectrum was
probably true to its reported temperature since the \keck\ iodine
atlas fitted the data very well, as described in the previous
section. Choosing a region with temperature-sensitive
lines, we found the best-fit temperature for the \keck\ cell's KPNO spectrum was
55~$\degree$C, although synthetic spectra ranging from 40-70~$\degree$C
all had a similar goodness of fit and they were hard to distinguish by eye
(column density and temperature are parameters with some degeneracy in the fitting).
We thus conclude that IodineSpec5 is reliable for
estimating temperatures for iodine FTS spectra, at least to an accuracy
of 5-10~$\degree$C.

\begin{figure}
\centering
\subfloat{\includegraphics[scale=0.5]{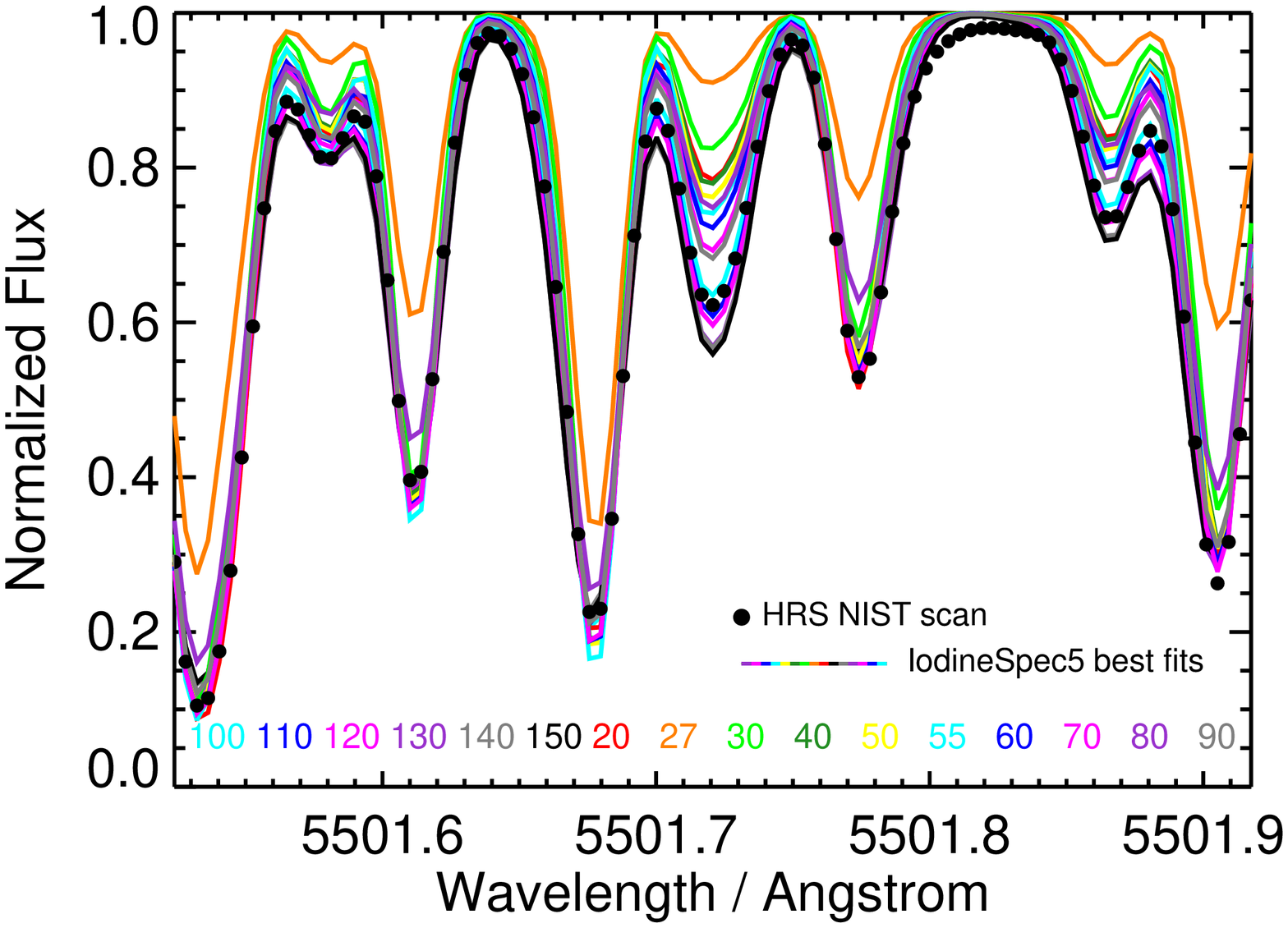}}\
\subfloat{\includegraphics[scale=0.5]{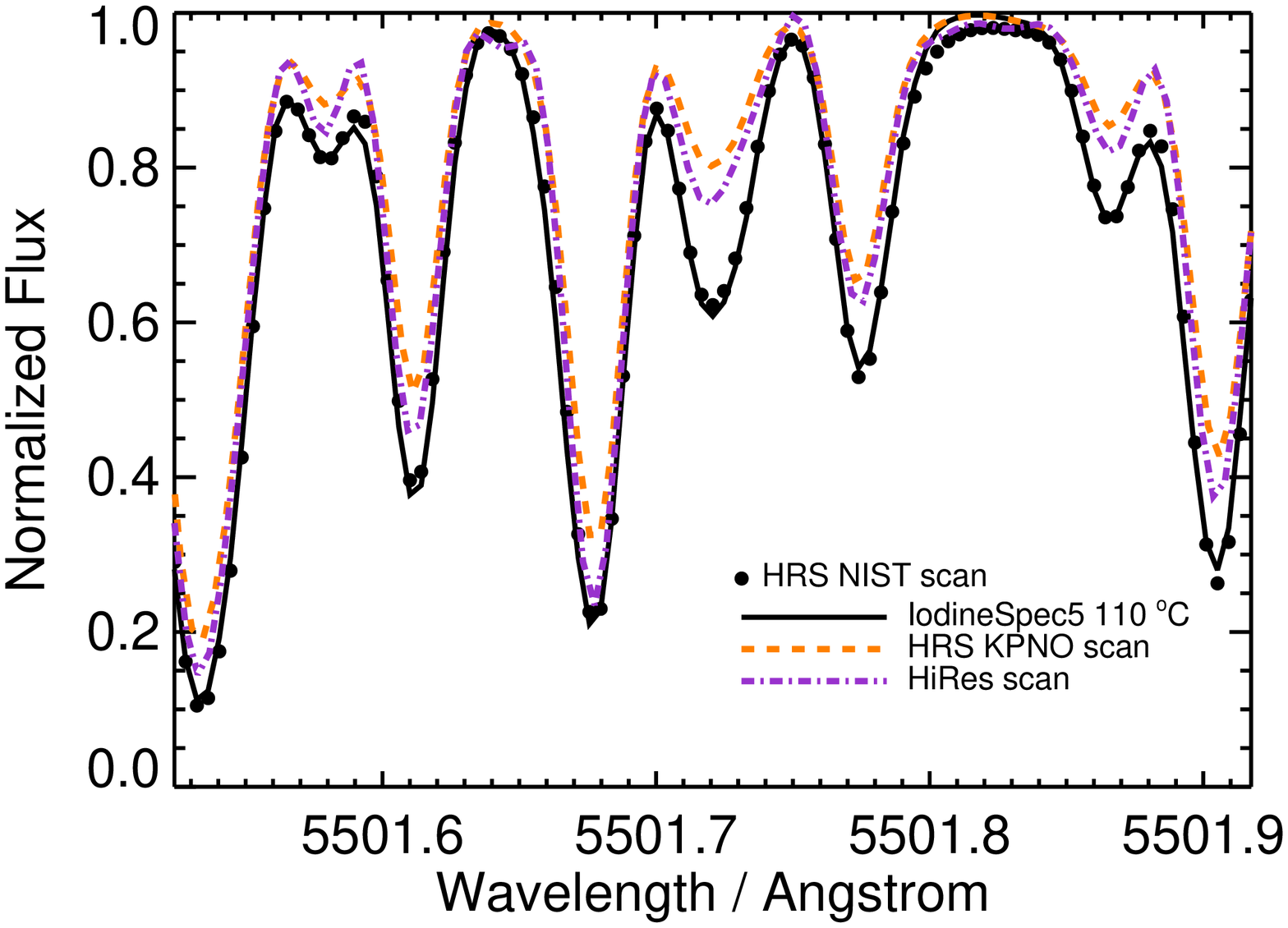}}
\caption{{\bf Top:} Fitting the \het\ NIST spectrum (black dots; when temperature was
set at 70~$\degree$C) with the best-fit IodineSpec5 synthetic iodine lines at
various temperatures (colored lines, with corresponding temperatures 
labeled in colored numbers). {\bf Bottom:} The \het\
NIST spectrum (black dots) over-plotted with the best-fit IodineSpec5 model
(black solid line; at 110~$\degree$C) and the KPNO FTS spectra for the \het\ cell (orange dashed) and the \keck\ 
cell (purple dot-dashed). All spectra in both panels are convolved down to a
resolution of 200,000 (roughly at the \keck\ cell's KPNO FTS spectral resolution) to
minimize the difference in line shapes due to IP differences between the FTS spectra (with a sinc function
IP) and also the synthetic spectrum (with only natural thermal broadening kernels).
\label{het:fig:nistfit}}
\end{figure}

We then fitted the NIST spectrum, which has the highest SNR and the highest
resolution among all FTS spectra and the TS12 spectra (we also had a rough
idea about its temperature from the visual comparison in 
Figure~\ref{het:fig:nisteyeball}). The results are shown in
Figure~\ref{het:fig:nistfit}, where the top panel shows the best-fit
models at different temperatures, and the bottom panel compares the
NIST spectrum with its best-fit model at 110~$\degree$C, the KPNO spectrum,
and the \keck\ cell's KPNO spectrum.

When we fitted the \het\ KPNO spectrum, the high degeneracy between
column density and temperature hindered us from getting an accurate
estimate for the temperature. Models in 40-80~$\degree$C appear to fit
equally well with varying column densities. However, only the fit at
$70\degree$C has the same column density as the best-fit value derived
from the NIST fit. We thus fixed the column
density in our fits for the \het\ cell's KPNO spectrum, and the best-fit
temperature came out to be $70\degree$C, as expected. Therefore, we conclude that during the NIST FTS spectrum,
the HRS cell was at a different gas temperature than the set temperature 
of $70\degree$C, and the cell was at the right temperature during the KPNO spectrum.

\begin{figure}
\centering
\subfloat{\includegraphics[scale=0.5]{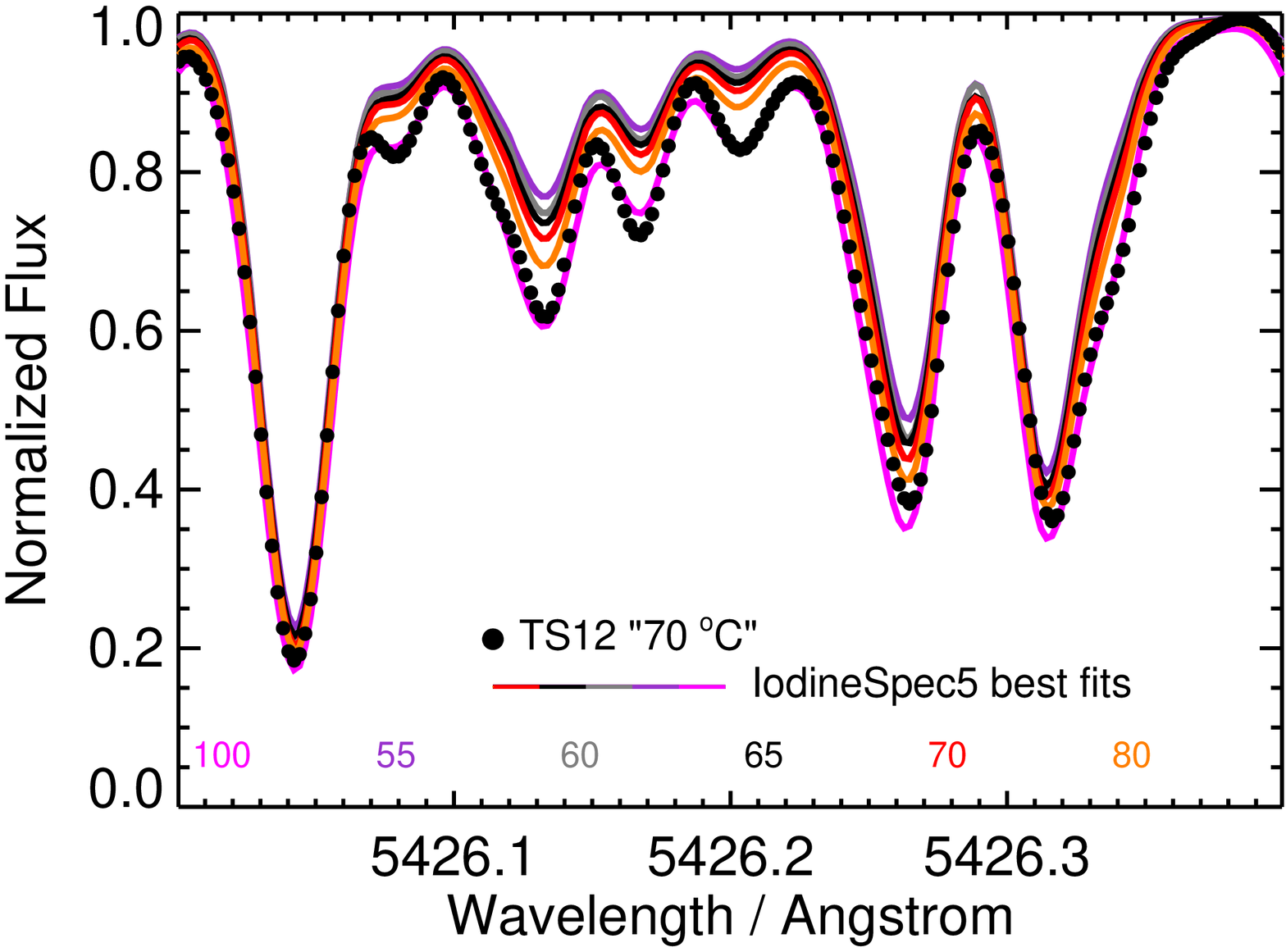}}\
\subfloat{\includegraphics[scale=0.5]{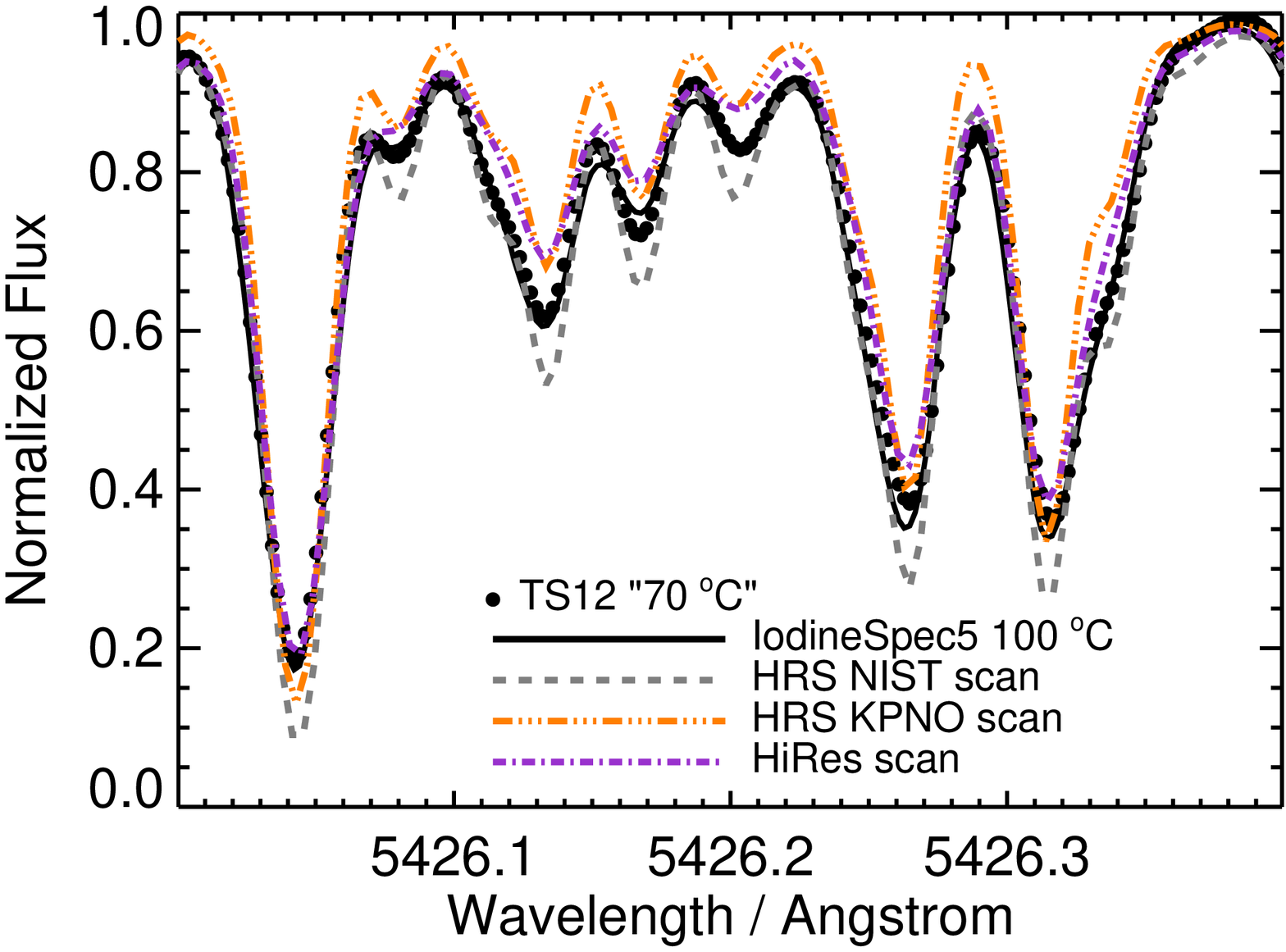}}
\caption{{\bf Top:} Fitting the TS12 spectrum (temperature set at
70~$\degree$C; black dots) with IodineSpec5 models, with fixed column
density derived from the best fits for the \het\ KPNO and NIST FTS spectra. 
Again, colored lines are best-fit IodineSpec5 models with different 
temperatures labeled in colored numbers. 
{\bf Bottom:} The best-fit temperature for the TS12 spectrum is
100~$\degree$C (black solid line). It is clearly at a lower temperature than
the NIST spectrum (gray dashed) but at a higher one than the KPNO spectrum
(orange triple-dot-dashed). For comparison, the HIRES cell's FTS spectrum is also 
plotted (purple dot-dashed). Again, all spectra in both panels were convolved
down to $R=$ 200,000.
\label{het:fig:ts12fit}}
\end{figure}

But how about the TS12 spectrum? Again, using the best-fit column
density derived from our best-fits for the NIST and KPNO FTS spectra, we estimated the
temperatures for the three TS12 spectra with temperatures reported to be set at 50, 60, and
70~$\degree$C. The best-fit temperature turns out to be 55~$\degree$C
for claimed 50~$\degree$C TS12 spectrum, 80~$\degree$C for the
60~$\degree$C one, and 100~$\degree$C for the 70~$\degree$C one. The
results from fitting the ``70~$\degree$C'' TS12 spectrum are illustrated in
Figure~\ref{het:fig:ts12fit}.

These findings on the \het\ cell temperatures could explain the fits to
the observed data. If the \het\ cell was kept at a higher temperature (e.g.,
$\sim 100\degree$C) instead of $70\degree$C when it was in active use
for precise RV calibration (despite what the temperature control
reported), or if the actual temperature of the gas in the cell changed
over time, then the KPNO FTS spectrum, which was done at $70\degree$C,
certainly cannot fit the observed data very well or provide precise
calibrations to measure RVs to the same level as \keck, which had a 
correct iodine atlas. The bottom panel of
Figure~\ref{het:fig:ts12fit} provided some clues for why the \keck\
cell's FTS spectrum provided the best fits to \het\ data
(Figure~\ref{fig:lampi2fit}) -- if the gas temperature was between 70
and 110~$\degree$C during actual HRS observations (and perhaps more often
closer to 70~$\degree$C), then an iodine atlas which has line depths in
between the KPNO (at 70~$\degree$C) and the NIST (at 110~$\degree$C) FTS spectra 
would provide a better fit than both, which the \keck\ cell's FTS spectrum happened to satisfy by coincidence.

Why was the temperature of the iodine gas in the cell higher than
70~$\degree$C as set by its temperature controller? We explore 
potential causes for the rise of HRS cell temperature in the next section.

\section{Cause for a Higher Iodine Gas Temperature}\label{sec:causetemp}

Throughout the FTS spectra at NIST and our TS12 observation, the readings
of the temperature probe on the \het\ iodine cell stayed at
70~$\degree$C. However, as discussed above, the temperature of
the iodine gas in the cell appears to be higher: 110~$\degree$C for the
NIST FTS spectrum and 100~$\degree$C for the TS12 spectrum with the
$70\degree$C setting.

One potential explanation for this temperature discrepancy is that the temperature probe was
malfunctioning. It does not report a true temperature but is biased
towards a lower reading than the actual temperature. In preparation for HRS's instrumental upgrade, the thermal enclosure and temperature control system was rebuilt since our TS12 experiment, so fortunately this old thermal system, functioning properly or not, would no longer affect any future HET's RV observations.

An alternative explanation is that the temperature probe was working,
but either there was a temperature gradient in the cell, or the gas
was at a higher temperature while the glass cell stayed cool. These
two alternative scenarios seem plausible, especially as an explanation
for the high temperature of the gas for the NIST FTS spectra. We noticed that the halogen lamp 
used for the NIST spectrum, which provided
the continuum emission shining through the iodine cell, was
exceptionally bright compared with other commonly used halogen 
lamps, probably in order to provide the high SNR needed for the high resolution FTS spectra. 
It was a lamp with a power of $1\ kW$ and the
room was considerably warmed up by this lamp. We discuss the
possibility of these two scenarios below through order-of-magnitude
estimates. The properties of the HRS iodine cell are as follows: it is a Pyrex glass cylinder cell, 70~mm in length and 50~mm in diameter with 6.35~mm thick windows. The cell is shorter than the typical 4~inch cell (such as the HIRES cell) in order to fit into the spectrograph, and as a result, it was run at a higher temperature of 70~$\degree$C than the HIRES cell to compensate for the shorter column length. We used a gas pressure similar to the HIRES cell, at 0.01~atm at its operating temperature (\citealt{butler1996};), which may not be accurate for the HRS cell but should suffice for our order-of-magnitude calculations.

\subsection{Temperature Gradient in the Glass}

We first explore if shining a very bright light, such as using the Xenon 
lamp at NIST,\footnote{It was a Newport 6269 Replacement Xenon Lamp (\url{https://www.newport.com/p/6269}).} 
at the HRS iodine cell would induce a
large enough temperature gradient in the cell glass
(Pyrex) to explain the difference between the gas temperature
(110~$\degree$C) and the temperature probe reading (70~$\degree$C,
assuming correct). The \het\ cell had one temperature probe taped 
at one of the side windows of the cell (the \keck\ cell has two 
probes: one in the middle within the heating wraps and one at the
window of the cell). A temperature gradient in the glass cell could mean
that the heated glass on one side can bring the gas inside to a higher
temperature, while the glass on the other side with the temperature
probe stays at $70\degree$C. To sustain a temperature gradient, there
needs to be a continuous heat input, which would be the Xe lamp
shining on the cell at NIST. According to the law of heat conduction, or
Fourier's law (in its 1-D differential form), the temperature gradient
can be estimated via: 
\beq
q = -k \frac{dT}{dx},
\eeq
where $q$ is the heat flux density in $W/m^2$, $k$ is the thermal
conductivity of the material in $W/m \cdot K$, and $dT/dx$ is the
temperature gradient in $K/m$.

The heat flux density coming from the Xe lamp can be derived by
\beq
q = I_{\rm lamp} \cdot \Delta \lambda \cdot f_{\rm absorption}, \label{eqn:q2}
\eeq 
where $I_{\rm lamp} = 200$ $mW/m^2/$nm is the irradiance of the Xe lamp used in the scan for the wavelength region of 400--700~nm at a distance of 0.5~m,\footnote{The irradiance spectrum of this Xe lamp can be found on its documentation in Figure 10 on Page 26 at \url{https://www.newport.com/medias/sys_master/images/images/hfb/hdf/8797196451870/Light-Sources.pdf}. This irradiance is an approximate number read off the plot, and the spectrum is fairly flat in the 400--700~nm region.} which is set by two colored glass filters (GG495 and BG40) that restrict the emission to the iodine region. Therefore $\Delta \lambda$ would be 300~nm. We assume a distance of 0.5~m here since we do not have the exact measurement of the setup at the NIST FTS scan, but this number should be close enough for the purpose of this estimate. $f_{\rm absorption}$ is the fraction of radiation absorbed by the glass.

To estimate $f_{\rm absorption}$, we need the absorption spectrum of cell material, Pyrex. Pyrex is largely transparent in the optical, but it has an absorption feature near 2.7 $\mu$m and it heavily absorbs UV light short of 300 nm and IR light longer than 5 $\mu$m.\footnote{The transmission spectrum of Pyrex can be in found in this website: \url{http://www.me.mtu.edu/~microweb/GRAPH/Laser/GLASS.JPG}, or can be easily found at several other places online documenting the properties of Pyrex.} Considering the filters used between the lamp and the iodine cell at the
NIST scan, $f_{\rm absorption}$ would mostly come from 400--700~nm region, where its average $f_{\rm absorption}$ is roughly 8\%. Therefore, the heat flux density coming from the Xe lamp would be $q=4.8~W/m^2$.

For a broader application of our estimate here, we also want to consider other common setups at astronomical observatories. Here we also estimate the heat flux density coming from a commonly used halogen lamp at a distance of 1~m with just a UV filter in between the lamp and the iodine cell. The heat flux density coming from the halogen lamp can be derived by
\beq 
q = P_{\rm lamp} \cdot f_{\rm beaming} \cdot f_{\rm absorption}
/4\pi D^2, \label{eqn:q}
\eeq where $P_{\rm lamp}=1\ kW$ is the power of the lamp,
$f_{\rm beaming}$ is the beaming factor of the lamp (as it is not
isotropically radiating towards all directions), $f_{\rm absorption}$
is the fraction of radiation absorbed by the glass, and $D$ is the
distance from the lamp to the cell. 

For a rough estimate, here we assume $f_{\rm beaming}=3$ (meaning the light from the lamp is filling a solid angle of $4\pi/3$) and $D=1$ m. 
To estimate $f_{\rm absorption}$, we assume the halogen lamp spectrum
is a black-body spectrum with peak around $1 \mu$m, which is typical
for this type of lamps. Considering the UV filter between the lamp and the iodine cell, $f_{\rm absorption}$ would mostly come from IR. This fraction can be estimated by numerically integrate the Planck function for the intervals $[2.5,3.0]\ \mu$m and $[5.0, \inf]\ \mu$m, and $f_{\rm absorption}$ turns out to be about 6\% of the total lamp emission. From these numbers, $q$ is estimated to be 14.3 $W/m^2$. Since this number is much larger than the heat flux density from the Xe lamp at NIST, we carry on our calculation using this number from the halogen lamp setup. As shown later, this heat flux density is not large enough to explain the temperature anomaly we saw, and therefore the same conclusion holds for the NIST setup using Xe lamp. 

Now we continue with our estimate for heat conduction in the iodine cell. The thermal conductivity ($k$) of Pyrex is a well known number, as it is a widely used and manufactured material, and it is about $k=1.1\ W/m\cdot K$.\footnote{From \url{
http://www.azom.com/article.aspx?ArticleID=4765}, for example.}

Plugging in $q=14.3~W/m^2$, $k$, and using $dx=7$ cm or 0.07~m (the length of the
iodine cell), $dT$ turns out to be about
1.0~$K$. Considering the fact that the light was shining on the window
of the glass cell and then being conducted down the thin cylinder
shell, and perhaps the distance between the lamp and the cell $D$ 
was smaller than 1~meter, $q$ may be boosted by a factor of a few
(assuming the thickness of the glass cylinder is a couple mm), up to 
perhaps 10~$K$. Therefore, we conclude that the temperature gradient is unlikely to be as large
as $40\ \degree$C, though not completely ruled out given our crude method of
estimation. In any case, this amount of heat may not be large enough to
heat the glass from $70\ \degree$C to $110\ \degree$C, or even sustain the
high temperature of the glass, because of its relatively high cooling rate estimated below. 

First, the amount of input energy from the lamp is probably too small to
cause a significant temperature rise in the glass. Using the specific
heat of Pyrex, $C_P = 0.75\ J/g\cdot K$,\footnote{From
\url{http://www.engineeringtoolbox.com/specific-heat-solids-d\_154.html}}
assuming the weight of the glass cell to be $M_{\rm glass} = 100\ g$,
and using an input power of $q \times \pi r^2$, where $r=0.025$~m is
the radius of the iodine cell window, the rate of the temperature rise
in Pyrex from heating of the lamp is:
\beq
dT/dt = q\cdot \pi r^2 / C_P M_{\rm glass} = 3.8 \times 10^{-4}\ K/s.
\eeq
Even assuming the glass does not cool, this means the temperature would only rise
for about 4 $K$ in an hour. 

Second, the thermal radiation from the glass alone would cool the
glass off fast enough. At $T=100\ \degree$C $= 375.15\ K$, the thermal
radiation has a power of $\sigma T^4 = 1123\ W/m^2$ -- much larger than
the input energy from the halogen lamp (recall that $q=14.3\ W/m^2$). 
Even considering that the glass was wrapped in thermal insulators and only its two windows were
exposed to the air, the thermal radiation would still be large enough to cool the glass down
fast enough and prevent continuous rising of the glass temperature (which would be too slow
anyway, as argued above).

Therefore, if the temperature probe was working, it was mostly likely that the
entire glass stayed at its reported temperature instead of having a
temperature gradient as large as $40\ \degree$C during the NIST FTS spectra.

\subsection{Heated Gas within a Cooler Glass}

Next, we investigate the temperature of the iodine gas inside the cell 
and whether it might be decoupled in temperature from its glass cell container. 
If the gas was heated up to a higher temperature without the glass cell being 
heated up considerably, then perhaps the temperature probe was working but it
was only reporting the temperature of the glass instead the gas. Below we argue 
that this is unlikely to be the case.

First, the small mean free path of the iodine molecule in the
cell suggests that they primarily thermalize with themselves and then
with the glass, which seems to support the hypothesis of the thermal 
decoupling between the gas and the glass, but we argue later with other 
evidence that is against this hypothesis. The mean free path of iodine 
molecules is estimated by using ideal gas assumption:
\beq
\lambda = (\sqrt{2} \sigma \cdot n)^{-1} = \frac{k_BT}{\sqrt{2}\pi d^2 P},
\eeq
where $\sigma$ is the interaction cross section of the molecule, $n$
is the number density of molecules, $k_B$ is the Boltzmann constant,
$d$ is the effective diameter of the molecule, and $P$ is the pressure
of the gas. The iodine gas in the cell has a pressure of about 0.01
atm or 1 kPa \citep{butler1996}, at a temperature of $110\ \degree$C or
385~$K$, with an effective diameter of about 5\AA\
\citep{JUHOLA1975437,topley1926size}. Using these numbers, the mean
free path is estimated to be 4.8 $\mu$m, much smaller than the size of
the glass cell. This means that the gas is not tenuous enough so that
their thermalization is completely dominated by the glass
enclosure. But more evidence is needed to support this hypothesis of thermal decoupling.

Second, assuming the gas was thermally decoupled from the glass, then 
what was the dominating heating source for the iodine gas? There were three
heating sources for the gas during the NIST FTS scan: (1) the heat from the
temperature enclosure heater conducted via the glass, (2) the heat from
the halogen lamp absorbed by the glass then conducted to the gas,
and (3) the direct radiation from the halogen lamp. For heat source (1), according to the
calculation in the previous subsection and our assumption here of a working temperature probe, 
the glass cell would have stayed at $70\ \degree$C, so
it was not heating up the gas to a higher temperature. 

For heat source (2), the additional heat received by the glass from the lamp,
we have worked out the amount to be 14.3 $W/m^2$ in the previous subsection. This is too small
to heat up the glass, but could this small amount of heat be conducted
to the iodine gas inside and heat it up? There are four outlets for
this received energy: radiative cooling of the glass, thermal
conduction to the rest of the glass cell, thermal conduction to the
air, and thermal conduction to the iodine gas within. The thermal
conductivity of the Pyrex glass is 1.14 $W/m\cdot K$, as mentioned
before, and the thermal conductivity of the air is about $2 \times 10
^{-2}\ W/m\cdot K$. Both are much larger than the thermal conductivity
of the iodine gas, which is on the order of $10^{-4}$-$10^{-3}\
W/m\cdot K$.\footnote{The estimates for the thermal conductivity of
iodine gas comes from Page 36 of \cite{vargaftik1993handbook}, which
gives the value of $\sim 4\times10^{-3}\ W/m\cdot K$ for a gas
pressure of about 0.2-0.5 atm. The estimate of $10^{-4}\ W/m\cdot K$
comes from theoretical values assuming ideal gas, $k=n\langle v \rangle\lambda
C_V/3N_A=k_B\sqrt{8RT/\pi M}/\pi d^2 \simeq 10^{-4}\ W/m\cdot K$,
where $M$ is the molar mass of iodine molecule.} This means that the
glass will primarily conduct along the glass cell itself, and then a
smaller fraction to the air, and then an even smaller fraction to the
iodine gas. Considering the factor of $100-1000$ difference in thermal
conductivity between glass/air and iodine gas, probably only a $< 1$\%
of the heat input $q=14.3\ W/m^2$ is being conducted to the iodine
gas, i.e., $< 0.14\ W/m^2$. This number would be even smaller if
thermal radiation of the glass is considered.

For heat source (3), the direct radiation from the halogen lamp, we estimate below that it
would be contributing a power of $q \sim 1.7\ W/m^2$ (or $3.3\ mW$, given 
the radius of the cell window is 2.5 cm) to the
iodine gas. This number is estimated by using equation \ref{eqn:q} for
$q$, but substituting $f_{\rm absorption}$ with $0.3 \times
0.024$, 0.3 being the approximate fraction of light the iodine molecules
absorbs in the optical window $[500,650]$~nm,\footnote{This was estimated from
a normalized iodine spectrum, so this could be a lower limit since no continuum absorption was taken into account.} and 0.024 is the fraction of energy
emitted by the halogen lamp in this wavelength range 
(again assuming its a blackbody as before). If the gas was
heated to above $70\ \degree$C, this would be the primary source of
heating. However, this amount of power is probably too small. Again,
the thermal radiation of any $100\ \degree$C material is $1123\ W/m^2$, as
calculated above, much larger than the $1.7 W/m^2$ input power. Plus
the iodine gas also cools via band emission
\citep{waser1947emission}. Therefore, it is very unlikely that the
iodine gas was actually heated up to $110\ \degree$C because of the
halogen lamp.

Additionally, although the thermalization between the glass and the
gas might be slow, it would not be at a longer timescale than
the FTS scan (about a few hours). The specific heat of Pyrex is
$C_P = 0.75\ J/g\cdot K$, and assuming the iodine gas is at
$T=100\ \degree$C and the glass is at $T=70\ \degree$C:
\beq
dT/dt = \frac{(\sigma T_{\rm gas}^4 - \sigma T_{\rm glass}^4)\cdot
  A}{C_P M_{\rm glass}} = 0.05\ K/s,
\eeq 
where $A$ is the surface area of the iodine cell, on the order of
$10^{-2}\ m^2$. This is probably a lower limit, because the iodine gas
would also emit band emission which will be absorbed by the glass (in
UV) and the glass would not cool very efficiently as it was wrapped in
thermal insulators. Therefore, if the gas was heated up to
$100\ \degree$C, then the glass would be heated up by the gas fast
enough to change the reading on the temperature probe on a timescale of minutes.

To summarize the findings above, it is highly unlikely that the glass
underneath the temperature probe was at $70\ \degree$C while the iodine
gas inside was at $110\ \degree$C. This means that a biased temperature
probe and/or a malfunctioning temperature controller is the most likely
explanation for high temperatures we found for the NIST FTS spectra and the 
TS12 spectra in Section~\ref{sec:fittemp}. One possible
scenarios is that, if the heater
of the temperature controller was turned on the entire time, the
temperature of the cell and the gas would have kept rising until they
had reached a thermal equilibrium, for example, with its thermal
radiation. If this were the case, then the power of the temperature controller heater
would be $1123 W/m^2 \times 10^{-2}\ m^2 = 11\ W$, which sounds plausible. This
could explain why all three NIST spectra, supposedly taken at 65, 70,
and 75~$\degree$C, all seemed to be at $110\ \degree$C instead. 

We conclude that the difference between the iodine atlas
and the iodine observations was a result of mismatched iodine gas temperatures, 
and this would be the most likely culprit behind \het's under-performance in terms of RV precision
compared with \keck. 

\section{Summary and Conclusion}\label{summary}

Iodine cell atlases are critical for obtaining precise RVs with iodine-calibrated high resolution spectrographs. In this paper, we investigated the reason behind the difference between the two iodine FTS atlases of the HET HRS iodine cell, and concluded that it was due to a temperature difference of the iodine gas caused by a malfunctioning thermal enclosure and controller. 

We have established a new method for validating iodine FTS atlases or performing quality checks of iodine cells in general --- using spectra taken by the ultra-high resolution arm, TS12, of the Tull Spectrograph One at McDonald Observatory. We have demonstrated that the TS12 spectra have similar quality to the FTS spectra, and in principle, the TS12 spectra could serve as the ``true solution" of the iodine spectrum in forward modeling if wavelength calibrated. We have also found that the software IodineSpec5 is very useful for determining the iodine gas temperature through fitting theoretically computed iodine absorption spectra to high SNR, high resolution iodine spectra, such as FTS spectra or the TS12 spectra. We also explored and ruled out two alternative explanations for the high temperature we found for the HRS iodine cell besides a malfunctioning temperature controller: a temperature gradient in the glass cell and thermal decoupling of the iodine gas and the glass cell.

Based on our work in this paper, we recommend that when diagnosing RV precision problems, it would be helpful to start with checking goodness of fit for the iodine calibration frames (lamp plus iodine, or O/B/A type star plus iodine observations\footnote{Observers should take care that stars used for this purpose truly have smooth continua from high rotational broadening.}). Ideally, the residuals of fits to these frames should be very close to a photon-limited precision. In addition, for iodine cells that would be exposed to extreme hot or cold weather, and especially short cells with a relatively large surface area at the windows, we recommend two temperature probes --- one at the body within the heating wraps and one at one of the windows near the light path.

When the validity of the iodine atlas is in question, one could resort to IodineSpec5 to fit for the temperature to see if it is consistent with the working temperature of the cell. Besides IodineSpec5, \cite{perdelwitz2018} has taken a suite of iodine FTS spectra under a series of temperatures from 23~$\degree$C to up to 66~$\degree$C. Ultra-high resolution echelle spectra taken through TS12 or using PEPSI (Potsdam Echelle Polarimetric and Spectroscopic Instrument; \citealt{pepsi2015}) on the Large Binocular Telescope can be helpful in diagnosing iodine atlas problems as well. 

In addition, given that the \keck\ iodine atlas is at a resolution around 250,000, it would be reasonable to speculate that PEPSI spectrum, upon wavelength calibration with a standard atlas and/or IodineSpec5, could be used as the iodine atlas in replacement for FTS scans as well. However, since \keck\ operates at a resolution of $\sim60,000$, more tests need to be done to validate if an iodine atlas at $R\sim$250,000 would be sufficient to model spectra at higher resolutions such as $R\sim$120,000. We have also briefly explored the possibility of using the spectra generated by IodineSpec5 (with the best-fit temperature and optical depth) as the iodine atlas to fit a selected 2\AA\ chunk (note: larger than the typical $\sim0.5$\AA\ areas we chose to present for temperature diagnoses in the figures) of spectrum from an iodine calibration frame taken by \het, and concluded that the IodineSpec5 spectrum is not accurate enough to serve as an atlas.

Given that the iodine atlas provides the first order ``ground truth" in precise RV work with iodine cells as calibrators, it is important to make sure that they accurately describe the state of the cell at its regular working stage. We recommend that, for iodine-calibrated precise RVs works, routine diagnostic data analysis on the iodine calibration spectra should be performed to track the quality and stability of the iodine cell as a calibrator.

\acknowledgments

We thank Anita Cochran and the astronomers who used the Cassegrain instruments at night during our TS12 runs (VIRUS-W for the first run and IGRINS for the second). We are deeply grateful for the help and support by the staff at McDonald Observatory --- our TS12 runs would not be successful without their devoted work. SXW is grateful for Iouli Gordon for his comments and for bringing to our attention the software package IodineSpec5. We greatly appreciate the work by Stephen Redman and Gillian Nave at NIST of performing the FTS scans of the HET/HRS cell as well as the data reduction. We thank Ming Zhao and Kimberly M.~S.~Cartier for conducting the TS12 observing run on the McDonald 2.7~m iodine cell and the HET/HRS cell and Ming Zhao's help on reducing the TS12 data. We also thank Gillian Nave, Larry Ramsey, and Suvrath Mahadevan for their helpful comments and suggestions. 

S.X.W.\ acknowledges support from NASA Earth and Space Science Graduate Fellowship (2014-2016). J.T.W.\ and S.X.W.\ acknowledge support from NSF AST-1211441. This work was also partially supported by funding from the Center for Exoplanets and Habitable Worlds, which is supported by the Pennsylvania State University, the Eberly College of Science, and the Pennsylvania Space Grant Consortium. 

This work herein has used observations obtained at the Hobby-Eberly Telescope at McDonald Observatory, and the HET partners include University of Texas at Austin, the Pennsylvania State University, the Ludwig Maximilians Universit\"at, and the Georg August Universit\"at.

The work herein has used observations obtained at the W. M. Keck Observatory, which is operated jointly by the University of California and the California Institute of Technology. The Keck Observatory was made possible by the generous financial support of the W.M. Keck Foundation. We wish to recognize and acknowledge the very significant cultural role and reverence that the summit of Mauna Kea has always had within the indigenous Hawaiian community. We are most fortunate to have the opportunity to conduct observations from this mountain. We also thank the California Planet Survey group for providing some of their iodine calibration data.

This work has made use of NASA's Astrophysics Data System Bibliographic Services. This research has made use of the SIMBAD database, operated at CDS, Strasbourg, France \citep{simbad}.

%

\vspace{5mm}
\facility{McDonald Observatories - Harlan J. Smith Telescope - Tull Spectrograph}
\facility{Keck Observatory - Keck Telescope - HIRES}
\facility{McDonald Observatories - Hobby-Eberly Telescope - HRS}


\software{IodineSpec5 \citep{iodinespec5}}

\bibliography{references}



\end{CJK*}

\end{document}